\pdfoutput=1
\documentclass[12pt]{article}
\usepackage{amsmath}
\usepackage{amssymb}
\usepackage[dvips]{graphics}
\usepackage{epsfig}
\usepackage{calc}
\usepackage{amsfonts}
\usepackage{graphicx}
\usepackage[ansinew]{inputenc}
\usepackage{color}

\newcommand{\appendixA}{\setcounter{equation}{0}
\def\theequation{\rm{A}.\arabic{equation}}\section}

\catcode`\@=11
\def\marginnote#1{}
\newcount\hour
\newcount\minute
\newtoks\amorpm
\hour=\time\divide\hour by60 \minute=\time{\multiply\hour by60
\global\advance\minute by-\hour}
\edef\standardtime{{\ifnum\hour<12 \global\amorpm={am}%
        \else\global\amorpm={pm}\advance\hour by-12 \fi
        \ifnum\hour=0 \hour=12 \fi
        \number\hour:\ifnum\minute<10 0\fi\number\minute\the\amorpm}}
\edef\militarytime{\number\hour:\ifnum\minute<10
0\fi\number\minute}
\def\draftlabel#1{{\@bsphack\if@filesw {\let\thepage\relax
   \xdef\@gtempa{\write\@auxout{\string
      \newlabel{#1}{{\@currentlabel}{\thepage}}}}}\@gtempa
   \if@nobreak \ifvmode\nobreak\fi\fi\fi\@esphack}
        \gdef\@eqnlabel{#1}}
\def\@eqnlabel{}
\def\@vacuum{}
\def\draftmarginnote#1{\marginpar{\raggedright\scriptsize\tt#1}}
\def\draft{\oddsidemargin -.5truein
        \def\@oddfoot{\sl preliminary draft \hfil
        \rm\thepage\hfil\sl\today\quad\militarytime}
        \let\@evenfoot\@oddfoot \overfullrule 3pt
        \let\label=\draftlabel
        \let\marginnote=\draftmarginnote
   \def\@eqnnum{(\theequation)\rlap{\kern\marginparsep\tt\@eqnlabel}%
\global\let\@eqnlabel\@vacuum}  }

\def\preprint{\twocolumn\sloppy\flushbottom\parindent 1em
        \leftmargini 2em\leftmarginv .5em\leftmarginvi .5em
        \oddsidemargin -.5in    \evensidemargin -.5in
        \columnsep 15mm \footheight 0pt
        \textwidth 250mmin      \topmargin  -.4in
        \headheight 12pt \topskip .4in
        \textheight 175mm
        \footskip 0pt
        \def\@oddhead{\thepage\hfil\addtocounter{page}{1}\thepage}
        \let\@evenhead\@oddhead \def\@oddfoot{} \def\@evenfoot{} }

\def\titlepage{\@restonecolfalse\if@twocolumn\@restonecoltrue\onecolumn
     \else \newpage \fi \thispagestyle{empty}\c@page\z@
        \def\thefootnote{\fnsymbol{footnote}} }

\def\endtitlepage{\if@restonecol\twocolumn \else  \fi
        \def\thefootnote{\arabic{footnote}} \setcounter{footnote}{0}}

\catcode`@=12 \relax
\def\bea{\begin{array}}
\def\eea{\end{array}}

\relax
\def\be{\begin{equation}}
\def\ee{\end{equation}}
\def\ba{\begin{eqnarray}}
\def\ea{\end{eqnarray}}
\def\del{\partial}
\def\d{{\rm d}}
\def\tr{\,{\rm tr}\,}
\def\tra{\,{\rm tr}_{\rm adj}\,}
\def\trR{\,{\rm tr}_R\,}

\def\r{\rho}

\def\b{\beta}
\def\g{\gamma}

\def\G{\Gamma}

\def\e{\epsilon}

\def\vf{\varphi}
\def\p{\psi}

\def\m{\mu}
\def\n{\nu}
\def\o{\omega}

\def\l{\lambda}
\def\L{\Lambda}
\def\s{\sigma}

\def\cM{{\cal M}}

\def\mt{{\tilde \mu}}

\def\psl{p\hskip-2.mm / \hskip0.5mm}
\def\Asl{A\hskip-2.mm / \hskip0.3mm}

\relax

\renewcommand{\theequation}{\thesection.\arabic{equation}}
\makeindex
\begin{document}
\topmargin-2.4cm
\begin{titlepage}
\begin{flushright}
LPTENS-07/18\\
April 2007
\end{flushright}
\vskip 2.5cm

\begin{center}{\Large\bf (Non) Gauge Invariance of Wilsonian Effective Actions }\\
\vspace{2mm} {\Large \bf in (Supersymmetric) Gauge Theories :}\\
\vspace{2mm} {\Large \bf A Critical Discussion}
\vskip 1.5cm 
{\bf Adel Bilal}\\
\vskip.3cm 
Laboratoire de Physique Th\'eorique,
\'Ecole Normale Sup\'erieure - CNRS\footnote{
unit\'e mixte du CNRS et de l'\'Ecole Normale Sup\'erieure 
associ\'ee \`a l'Universit\'e Paris 6 Pierre et Marie Curie}
\\
24 rue Lhomond, 75231 Paris Cedex 05, France

\end{center}
\vskip .5cm

\begin{center}
{\bf Abstract}
\end{center}
\begin{quote}
We give a detailed critical discussion of the properties of Wilsonian effective actions~$\G_\m$, defined by integrating out all modes above a given scale $\m$. In particular, we provide a precise and relatively convenient prescription how to implement the infrared cutoff $\m$ in any loop integral that is manifestly Lorentz invariant and also preserves  global linear symmetries such as e.g. supersymmetry. We discuss the issue of gauge invariance of effective actions in general and in particular when using background field gauge. Our prescription for the IR cutoff (as any such prescription) breaks the gauge symmetry. Using our prescription, we have explicitly computed, at one loop, many terms of the Wilsonian effective action for general gauge theories, involving bosonic and fermionic matter fields of arbitrary masses and in arbitrary representations, exhibiting  the non-gauge invariant (as well as the gauge invariant) terms. However, for supersymmetric gauge theories all non-gauge invariant terms cancel within each supermultiplet. This is strong evidence that in supersymmetric gauge theories this indeed defines a Lorentz, susy and gauge invariant Wilsonian  effective action. As a byproduct, we obtain the explicit one-loop Wilsonian couplings for all higher-derivative terms $\sim F D^{2n} F$ in the effective action of arbitrary supersymmetric gauge theories.
\end{quote}
\end{titlepage}
\setcounter{footnote}{0} \setcounter{page}{0}
\setlength{\baselineskip}{.6cm}
\newpage
\begin{small}
\tableofcontents
\end{small}

\section{Introduction}
\setcounter{equation}{0}

The notion of an effective action plays a most important role in modern quantum field theory. While it had long been believed that a basic criterion for any quantum field theory is its renormalizability, over the years it has become increasingly clear that our preferred renormalizable theories are just to be considered as the infrared limits of some more fundamental theories. At any finite energy scale one should actually include higher dimension operators in the action and view these theories as effective field theories described by some effective action.

There are many quite different objects going under the name of effective action. The common  feature is that they somehow describe the effective behavior of certain fields at low energy without having to worry in detail about the high energy physics that already has been "integrated out". Specifically, one may distinguish a set of heavy and a set of light fields and completely integrate out the heavy ones, obtaining an effective action $S_{\rm eff}$ for the light ones only.  Another notion of effective action is that of the generating functional $\G$ of one-particle irreducible (1PI) diagrams (proper vertices) where one already has computed all the loop-diagrams. A somewhat intermediate notion is that of Wilsonian effective action $\G_\m$ where for all fields one only integrates out the high momentum/high energy modes  above some scale $\m$. For loop diagrams this means that all loop-momenta are only integrated down to some infrared cutoff $\m$. When this Wilsonian effective action is used to compute correlation functions one only needs to do the remaining integrations over the low momentum/energy modes, i.e. perform loop integrals now with a UV cutoff $\L$ equal to $\m$. This last property is used to define the Wilsonian action in the context of the exact renormalization group (ERG) \cite{Pol}, where it is an effective action with a UV cutoff $\L$ obeying a certain flow equation that guarantees that the correlation functions do not depend on the UV cutoff $\L$. 

In the presence of massless fields, 
the 1PI effective action has infrared singularities, i.e. is non-analytic at zero momentum. On the other hand, the Wilsonian effective action allows an expansion  in powers of the momenta divided by $\m$ and thus is an (infinite) sum of local terms. It is this locality of the Wilsonian effective action that plays an important role in many places.

In supersymmetric theories there are important non-renormalization theorems for the superpotenial, or more generally for the $F$-terms of the action. This has been shown in perturbation theory using the powerful supergraph techniques \cite{supergraphnonren}. An alternative very elegant proof of these non-renormalization theorems was given by Seiberg \cite{SeibergNR} just based on the symmetries and holomorphy of the F-terms in the {\it Wilsonian} effective action. The proof deals with the Wilsonian action since locality is crucial in order to separate $D$ and $F$-terms.\footnote{
Similarly, the proof of non-renormalization using supergraphs could in principle be invalidated by infrared divergences.
} 
The same symmetry arguments actually also constrain the non-perturbative corrections to the $F$-terms.
It is most important for the proof, and usually assumed to be true, that the Wilsonian effective action is supersymmetric and Lorentz invariant. In a supersymmetric gauge theory it should also be gauge invariant. However, these properties are by no means obvious.

With this motivation in mind, in this note we would like to discuss these questions in some detail: 
how exactly do we define the Wilsonian effective action with the infrared cutoff $\m$? how do we make sure this definition and the introduction of $\m$ is Lorentz invariant and does not break gauge invariance or supersymmetry? 
	
In section 2, after recalling some important issues about 1PI effective actions, we will provide a detailed discussion of Wilsonian effective actions $\G_\m$. In this note, we will define the Wilsonian action by starting from a ``microscopic" theory and really integrate out all modes above a given scale $\m$. In particular, we will give a precise (and relatively convenient) prescription how to implement the finite {\it infrared} cutoff $\m$ for any loop diagram that is Lorentz invariant and respects the various linear global symmetries, like e.g. supersymmetry.  We will not use the flow equations of the ERG which are different in spirit. We discuss a simple one-loop example in scalar $\vf^4$ theory, as well as a two-dimensional example of chiral fermions coupled to an abelian gauge field, where one can explicitly see the transition in the Wilsonian effective action $\G_\m$ from a sum of local terms to a non-local expression as the ratio of momentum $p$ and $\m$ is varied from ${p\over \m }\ll 1$ to ${p\over\m}\gg 1$. The remainder of this section deals with the issue of gauge invariance of the effective actions. Here, we also discuss various approaches in the existing literature that are mainly concerned with the possibility to introduce invariant ultraviolet regularizations in the exact renormalization group and the corresponding flow equations. We further discuss the role of using background field gauge and the manifestation in the Wilsonian action of possible anomalies.

In section 3, using our prescription for the IR cutoff, we explicitly compute many one-loop terms of the Wilsonian effective action for general gauge theories involving bosonic and fermionic matter fields of arbitrary masses and in arbitrary representations. We find that the presence of the finite infrared cutoff $\m$ explicitly breaks gauge invariance, as expected, and the Wilsonian effective action for a generic gauge theory contains infinitely many non-gauge invariant terms. (Nevertheless, it will be evident that the physical correlation functions computed from this Wilsonian effective action do satisfy the Ward identities.) 
However, we will also show, at least for those (infinitely many) terms of the Wilsonian effective action we explicitly computed, that in a supersymmetric gauge theory, when adding the contributions of all fields within any ${\cal N}=1$ supermultiplet, the non-gauge invariant terms precisely cancel. We argue that this is strong evidence that in a supersymmetric gauge theory one can indeed introduce the infrared cutoff $\m$ and still have a Lorentz, susy and gauge invariant Wilsonian effective action {\it at any finite scale} $\m$. We use our results to explicitly give the one-loop Wilsonian couplings for all higher-derivative terms $\sim F D^{2n} F$ in the Wilsonian effective action for arbitrary supersymmetric gauge theories. 

In the appendix, we discuss in more detail how to implement the infrared cutoff $\m$ for arbitrary $L$-loop diagrams. To illustrate the procedure, we present a complete two-loop calculation in scalar $\vf^4$ theory of the Wilsonian $\G_\m^{(2)}$. Although there are a few subtleties not present at one loop, in the end we will obtain a very explicit result. 

\section{The Wilsonian effective action}

\setcounter{equation}{0}

\subsection{The 1PI effective action}

To compute correlation functions in any quantum field theory it is most convenient to first obtain the effective action $\G[\vf]$ which is the generating functional of one-particle irreducible (1PI) diagrams (proper vertices). As is well-known, within perturbation theory one can then obtain all diagrams contributing to a given correlation function by summing all {\it tree} diagrams made up with the effective vertices (which are 1PI) and full propagators as given by $\G$. In this sense, $\G$ already contains all effects from loops, and actually also includes contributions beyond perturbation theory. In particular, the whole issue of renormalization must be settled when computing the effective action $\G$. Also, all symmetries of the quantum theory are coded in $\G$. In particular, if the regulated functional integral measure does preserve any {\it linear} symmetry of the classical action (including possible gauge fixing terms), i.e. if these symmetries are non-anomalous, then $\G$ also is invariant under these same symmetries. This is usually expressed by Ward or Slavnov-Taylor identities. 

The issue of (non-abelian) gauge symmetries is more complicated since one has to add to the classical Lagrangian ${\cal L}_{\rm cl}$ a gauge fixing term ${\cal L}_{\rm gf}$ and a corresponding ghosts term ${\cal L}_{\rm gh}$ which of course break the gauge symmetry. The gauge symmetry then is replaced by the BRST symmetry of the complete Lagrangian ${\cal L}$ which is the sum of the three terms. The BRST symmetry acts nonlinearly  and hence the effective action has no reason to be BRST invariant, and much less gauge invariant. Instead, one can show that the effective action obeys the Zinn-Justin equation \cite{ZJ} or in more modern terms the Batalin-Vilkoviski (quantum) master equation \cite{BV}, which severely constrains the possible counterterms to be BRST invariant. 

An alternative approach consists in employing the so-called background field method which explicitly introduces a background gauge field and computes $\G[A]$ from $\int {\cal L}[A+A']$ by treating $A$ as a gauge field and $A'$ as transforming as a matter field in the adjoint representation. One can then introduce a gauge fixing for $A'$ such that the effective action $\G[A]$ still is manifestly invariant under the gauge transformations of $A$. Technically, integrating over $A'$ necessitates the knowledge of the $A'$ propagators and vertices in the presence of the $A$ background fields. They can be expanded in powers of $A$ reproducing the usual diagrammatic expansion with internal $A'$-lines and external $A$-lines (without propagators). It clearly provides a method, at least in principle, to define a {\it gauge invariant 1PI effective action} $\G[A]$. 

In supersymmetric gauge theories one can use a somewhat modified version of the background field method directly in superspace. The necessary modification is due to the fact that the vector superfield $V$ which contains the gauge and gaugino fields transforms in a complicated way under the gauge symmetry and a linear split of the form $V+V'$ is not appropriate. This is a slight complication only and this superspace background field method is well-known \cite{supergraphnonren} (see also \cite{West}). It guarantees manifest gauge and susy invariance of the effective action.  Note that for ${\cal N}=2$ extended supersymmetry the appropriate superspace is harmonic superspace and in this case there exists also a specific background field method which is even somewhat simpler \cite{BB}. In  any case, we can define a susy and gauge-invariant 1PI effective action using these methods.

In general, the 1PI effective action $\G[\vf]$ is a complicated {\it non-local} functional of the fields $\vf$. These non-localities are due to the momentum flow through the propagators in the loops. As long as no massless fields are present one can always expand these non-local terms in powers of (external) momenta over masses, resulting in a sum of local terms, although with arbitrarily many derivatives.
If the theory contains massless fields the proper vertices exhibit singularities at zero momenta and such an expansion is not possible. One can trace the origin of these singularities as coming from the region of small loop momenta. To illustrate this, consider the one-loop contribution to $\G$ in scalar $\vf^4$-theory with mass $m$. At one-loop, the two-point function only gets a momentum-independent constant contribution, while the four-point vertex function\footnote{
Our conventions are as follows:  Our signature is $(-,+,+,+)$. We let $\vf(x)=\int {{\rm d}^4 q\over (2\pi)^4} e^{-iqx} \tilde\vf(q)$ and then $\G=\sum_n {1\over n!} \int {{\rm d}^4 p_1\over (2\pi)^4} \ldots \int {{\rm d}^4 p_n\over (2\pi)^4}\, (2\pi)^4 \delta^{(4)} (\sum p_i)\ \G^{(n)}(p_i)\, \tilde\vf(p_1) \ldots \tilde\vf(p_n)$.  In particular, we have $\G^{(2)}\vert_{\rm tree}(p)=-(p^2+m^2)$ and $\G^{(4)}\vert_{\rm tree}(p_i)=-g$.
} 
$\G^{(4)}$ gets three contributions 
\be\label{G41PI}
\G^{(4)}_{\rm 1-loop}(p_i) = - {g^2\over  2} 
\left[J^{(4)}(-s) + J^{(4)}(-t) + J^{(4)}(-u)\right] \ ,
\ee
where (cf. the left part of Fig. \ref{4pointfct})
\be\label{G4}
J^{(4)}(P^2)={1\over (4\pi)^2}\left( c+\int_0^1 {\rm d}x\ \log[m^2+x(1-x)P^2]\right) \ .
\ee 
Here $s$, $t$ and $u$ are the usual squares of sums of two external momenta $s=-(p_1+p_2)^2$, etc, $x$ is a Feynman parameter and $c$ is constant\footnote{We consider the $J^{(4)}$ to be the one-loop expressions to which one still must add the contributions from the counterterms to make them finite.
Also note that in dimensional regularization the coupling constant is not dimensionless and one usually writes it as $m^\e g$ or $\m_0^\e\, g$ with some scale $\m_0$, so that the argument of the logarithm in (\ref{G4}) becomes dimensionless. However, for the purpose of comparing with the corresponding Wilsonian expression below (which involves an adjustable scale $\m$) it is preferable not to do so.
} 
(in dimensional regularization e.g. with $d=4-\e$ one has $c=-{2\over \e} +\g-\log 4\pi$).
The function $J^{(4)}(P^2)$ exhibits the usual unitary cuts 
for $-P^2\ge 4 m^2$. Nevertheless, in a massive theory, for $|P^2|< 4 m^2$ we can expand this in powers of $P^2/m^2$. After Fourier transforming, the corresponding contribution to $\G$ then is an (infinite) sum of terms that are {\it local}, i.e. involving a single integral over space-time $\int {\rm d}^4 x\ldots$, each one containing more and more derivatives. Of course, the effective action does not only contain 4-point vertices, but - a priori - all interactions that are consistent with the symmetries. For example, there is a six-point vertex function $\G^{(6)}$ which gets contributions from a one-loop triangle diagram 
\be\label{G6}
\G^{(6)}_{\rm 1-loop}(p_i)\sim
\sum_{\rm permutations}
\int_0^1 {\rm d}x \int_0^{1-x} {\rm d}y\ [m^2+x(1-x)P_1^2 +y(1-y)P_2^2 +2xy P_1\cdot P_2]^{-1} \ ,
\ee 
where each $P_i$ is the sum of the two external momenta flowing into the triangle diagram at the $i^{\rm th}$ vertex. Again, one can expand in $P_i\cdot P_j/m^2$ for small enough $P_i$ obtaining a sum of local contributions to $\G$. Obviously, this is no longer true if $m=0$: typically in a theory containing massless fields, the $\G^{(n)}(p_i)$ have branch cut singularities starting at zero momenta.

\subsection{Defining the Wilsonian effective action}

Since the singularities of $\G[\vf]$ in the presence of massless fields are due to the regions of small loop momenta, one way of avoiding them is to impose an IR cutoff in loop diagrams. This is exactly what one does when computing the {\it Wilsonian} effective action $\G_\m[\vf]$. It is computed just like $\G$ but with the restriction that all loop momenta are only integrated down to some (large) ``IR-cutoff" $\m$. This implies that, even for $m=0$, the Wilsonian effective action is {\it local} in the following sense:
As long as all external momenta are well below the scale $\m$ one can safely expand $\G_\m[\vf]$ in powers of momenta divided by  $\m$ obtaining an effective action that is an (infinite) sum of {\it local} terms.
This is a most important difference with the 1PI  effective action and one of the main reasons certain statements can be made about the Wilsonian action and not about the 1PI action.
Note that the whole issue of UV-divergences and renomalization has to be dealt with when computing the Wilsonian effective action - just as for the 1PI action.

More generally, one may want to ``integrate out" all high energy or high momentum modes.
In particular in situations with a hierarchy of masses where heavy particles have masses $M_i\gg\m$ and light particles have masses $m_a\ll\m$, one could as well completely integrate out the heavy fields\footnote{
A recent review of effective actions obtained by integrating out heavy fields is e.g. \cite{Burgess}.} 
and apply the IR cutoff $\m$ only to loops with light fields.
In practice though, this can be cumbersome as a loop could involve both light and heavy particles, and we will stick to the prescription that $\m$ is an IR cutoff for all loops. For loops involving heavy particles the difference between both prescriptions clearly is suppressed by a factor $\m^2/M^2$, as one can also check on our explicit examples below.

\begin{figure}[h]
\centering
\includegraphics[width=0.7\textwidth]{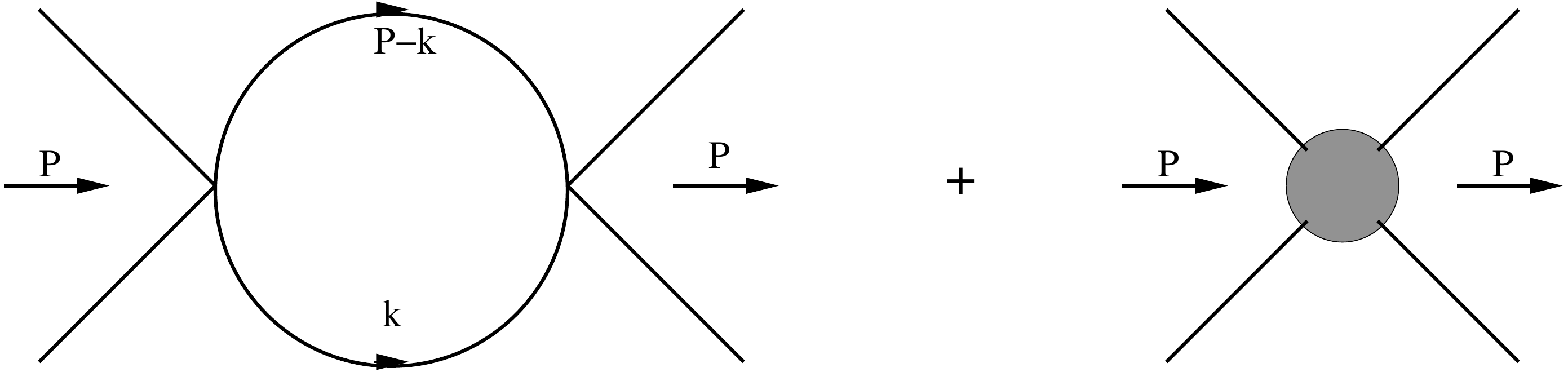}\\
\caption[]{Two ${\cal O}(g^2)$ contributions to a four-point function for scalars with a quartic interaction when using the Wilsonian effective action $\G_\m$. The loop diagram on the left involves two ${\cal O}(g)$ vertices and is to be computed with a UV cutoff $\m$, while the tree diagram on the right involves the ${\cal O}(g^2)$ vertex of $\G_\m$.} \label{4pointfct}
\end{figure}
The Wilsonian  $\G_\m$ is to be used as an effective action to compute
correlation functions. Since $\G_\m[\vf]$ only takes into account loop momenta above $\m$, one must now add tree diagrams {\it and loop diagrams} using the vertices and propagators from $\G_\m[\vf]$ and integrate the loop momenta from 0 to $\m$ which then serves as the {\it UV-cutoff}.\footnote{
For this reason, the authors of ref. \cite{SV} refer to the 1PI action $\G$ as a $c$-number expression and to the Wilsonian effective action $\G_\m$ as an operator expression.
}, see Fig. \ref{4pointfct}.
This reconstitutes the full integration range for the loop momenta.
It is this additional integration over the ``low momenta" that reproduces the IR singularities of the 1PI action $\G$. In this two-step procedure - first compute $\G_\m$ and then use it to compute correlation functions - $\m$ is only an arbitrary intermediate scale (it is a priori {\it not} the scale at which renomalization conditions are imposed), which should not affect the final answer for the correlation functions and drop out. 

This cancellation of the $\m$ dependence in the correlation functions is an important point. It is obviously true if one is only concerned with the one-loop approximation where there is a single loop momentum to be integrated, provided one uses the same implementation for the UV-cutoff as one used for the IR cutoff when computing $\G_\m$. However, for multi-loop diagrams with two (or more) loops sharing a common momentum, this can become quite tricky. (The problem is somewhat similar to the usual difficulty associated with overlapping UV divergences.)
At any rate, in order for the $\m$ dependence to cancel one must have a precise universal prescription of how to implement the IR cutoff $\m$ when computing $\G_\m[\vf]$ and then use the {\it same} prescription  for the UV cutoff $\m$ when using $\G_\m[\vf]$ as the action to compute correlators. Before turning to this issue, let us mention, however, that one can turn this argument around and {\it define} the Wilsonian effective action as containing all possible local interactions with $\m$-dependent coupling constants $g_n(\m)$ (and field normalization factors)  such that
\vskip8.mm

\begin{itemize}
\item when computing correlation functions with a UV-cutoff $\m$ all $\m$-dependence drops out, 
\item when $\m$ equals the UV-cutoff $\L_0$ (in which case there are no loop integrals left) the couplings $g_n(\m)$ equal the bare coupling constants of the classical action (typically with only finitely many being non-zero). 
\end{itemize}
Defining $\G_\m$ this way one does not really have to worry about complicated overlapping loops: they are still troublesome to evaluate in practice, but we don't have to worry about them in principle.
Also, this way of defining $\G_\m$ obviously is not restricted to perturbation theory or a diagrammatic expansion. It is the basis for the so-called exact renormalization group (ERG) \cite{Pol}, where the $\m$-independence of the correlators is the content of the flow equations.  Of course, in many theories and in particular in gauge theories, we do not want to use an explicit UV cutoff $\L_0$, and then it is not so clear how to implement the second requirement.\footnote{
There exist gauge invariant UV ``cutoffs" based on covariant higher derivative terms $\sim \left({ D^2\over \L_0^2}\right)^n$ added to the action \cite{Warr}. We will discuss them further in section 2.6 below.
} 
Also, this way of defining $\G_\m$ is somewhat less intuitive. For these reasons, we will {\it not} define $\G_\m$ this way, but instead keep with the first definition of explicitly integrating out all loop momenta above $\m$.

\subsection{Explicit realization of the infrared cutoff}

Let us now turn to the question of giving a precise explicit prescription how to implement the infrared cutoff $\m$ on the loop momenta. A basic criterion is that it should not break Lorentz invariance. It also must be independent of the way we label the loop momenta, i.e. it should be insensitive to shifts of the loop momenta.
Of course, it should also, as much as possible, preserve all other symmetries of the classical action.

First note that the problem is more complicated than the usual one of UV-regulating divergent diagrams. Indeed, many Lorentz-invariant ways are known to UV regulate diagrams with an explicit cutoff $\L$ (e.g. by working with modified propagators). One does not have to be too specific since, in the end, one takes $\L\to\infty$. Concerning the IR cutoff $\m$, however, we want to keep $\m$ finite and make sure that one can implement the same prescription for the UV cutoff $\m$ when using the action $\G_\m$ to compute correlators so that one reconstitutes the full momentum integrations (see Fig. \ref{4pointfct}) and the $\m$ dependence really cancels. The required cancellation of the $\m$ dependence excludes simple modifications of the propagators (like adding a mass term $\m^2$) to implement the IR cutoff.\footnote{
Note however, that for the purpose of studying the renormalization group flow one is essentially only concerned with infinitesimal {\it changes} of the IR cutoff to establish the flow equations. This allows for more flexibility in the choice of IR cutoff, like adding a term $\sim \int {\rm d}^4 q\ \tilde\vf(q) R_\m(q) \tilde\vf(-q)$ to the classical action \cite{Wetterich} with e.g. $R_\m(q)= q^2 (e^{q^2/\m^2}-1)^{-1}$ that modifies the propagators by a momentum dependent term giving effectively a mass to the low momentum modes $q^2\lesssim \m^2$ and not modifying the high momentum modes $q^2\gg\m^2$.
} 
We will now discuss two different ways how to separate the low and high momentum modes. They will provide explicit IR, resp. UV cutoffs which are manifestly Lorentz (and also susy) invariant.

\begin{figure}[h]
\centering
\includegraphics[width=0.7\textwidth]{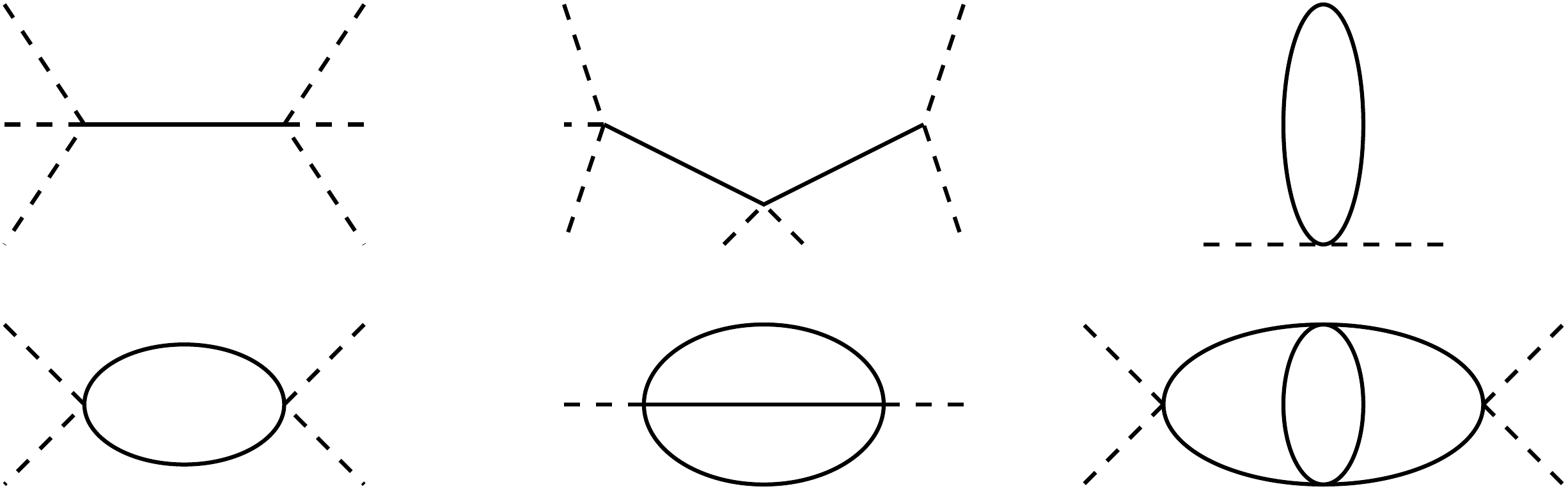}\\
\caption[]{Shown are a few contributions to $\G_\m[\vf_-]$ generated when integrating out $\vf_+$ in a theory with quartic interactions. The solid lines represent $\vf_+$ propagators, while the dashed ones represent external $\vf_-$.} \label{phipmdiag}
\end{figure}

One explicit way to separate high and low momentum modes (see e.g. ref.\cite{Peskin})  is to separate the Fourier expansion of each field $\vf$ (bosonic or fermionic) into two parts $\vf_+(p) + \vf_-(p)$ with $\vf_-(p)$ vanishing for $p_E^2 >\m^2$ and $\vf_+(p)$ vanishing for $p_E^2\le\m^2$, $p_E$ being the Euclidean momentum. Since the free action is diagonal in momenta the propagators do not mix $\vf_-$ and $\vf_+$. This splitting obviously is Lorentz invariant and also respects the global linear symmetries like e.g. supersymmetry: the off-shell algebra for {\it global} supersymmetry is linear in the fields and their derivatives and does not involve any explicit functions of space-time. Hence it commutes with the action of the projectors on low or high momentum modes, and the decomposition is susy invariant. Clearly, the same applies to any other global linear symmetry. One may then explicitly do the functional integral over the high momentum modes $\vf_+$:
\be\label{lhmodesint}
e^{i\G_\m[\vf_-]}=\int [D\vf_+] e^{i S[\vf_-,\vf_+]} \ .
\ee
This can be evaluated, at least in principle, order by order in an expansion in $\vf_+$-loops. Indeed, the $\vf_-$ play the role of external sources and only $\vf_+$ propagators ever appear in the expansion. Note that the expansion can also contain tree diagrams with $\vf_+$ propagators, see Fig. \ref{phipmdiag}. Obviously, such tree diagrams arise if several low momenta of the external $\vf_-$ meeting at a vertex add up to produce a high momentum of a $\vf_+$.

Obviously, for gauge theories, this separation into low and high momentum modes does not respect the gauge symmetry which is a local symmetry and hence non-diagonal in momentum: any (non-constant) gauge transformation like $\delta \phi(x)=i[\e(x),\phi(x)]$ will mix $\phi_+$ and $\phi_-$. Also, for practical calculations, this separation of modes quickly becomes very cumbersome.

Let us now describe an alternative method that (although not respecting gauge symmetry either) is more convenient for practical computations. It does not involve as sharp a momentum separation for each field as the method above, but instead separates {\it loop} momenta into two regions: larger and smaller than $\m$. Since one can always shift the loop-momenta one needs to give a specific prescription to avoid any ambiguities. To simplify the discussion, here we will only give the prescription for one-loop integrals. The generalization to any $L$-loop integral is relatively straightforward and will be given in the appendix, where we also do an explicit two-loop computation to illustrate this prescription.

One proceeds in the following way: First evaluate all tensor and gamma matrix algebra and introduce Feynman parameters $x_a,\ a=1,\ldots r$.  Then any one-loop diagram depending on $n$ external momenta $p_s$ and having $r$ propagators takes the form
\be\label{Ioneloop}
I(p_s) 
= (r-1)!\, \Big( \prod_{a=1}^r \int_0^1 {\rm d} x_a \Big)\
\delta\Big(\sum_{a=1}^r x_a -1\Big) 
\int {{\rm d}^4 k\over (2\pi)^4} \ {\cal I} \ ,
\ee
where the integrand is
\be\label{oneloopintegrand}
{\cal I}=Q(k,p_s) \left[ 
k^2 + 2 k\cdot {\cal P}(x_a,p_s) + C(x_a,p_s,m_a)\right]^{-r}\ . 
\ee
Here $Q$ is some polynomial in the momenta resulting from doing all the relevant spinor and tensor algebra, and it transforms in the appropriate representation of the Lorentz group. The bracket $[\ldots]$ is a Lorentz scalar provided we also transform the loop momentum $k^\n$. Then to regulate any UV divergences we use dimensional regularization. Actually we have two options: Usually, in dimensional regularization, one starts with fully $d$-dimensional Feynman rules and then the polynomial $Q(k,p_s)$ results from doing the tensor and $\g$-matrix algebra in $d$ dimensions. Alternatively, as is standard in supersymmetric theories, one can first do all tensor and $\g$-matrix algebra in four dimensions and only then do the dimensional regularization of the integrals. This latter procedure is known as dimensional reduction \cite{Siegel}. For the purpose of implementing the IR cutoff $\m$ one can use either version, as long as one does so coherently throughout.

The loop integral then is convergent and we can shift the loop momentum from $k$ to $k'=k+{\cal P}(x_a,p_s)$. This allows us to put the one-loop integral into the following {\it standard form}:
\be\label{standardint}
\int {{\rm d}^dk\over (2\pi)^d}\ {\cal I} =  
\int {{\rm d}^dk'\over (2\pi)^d}\,
Q(k'-{\cal P},p_s) \left[ {k'}^2 + C -{\cal P}^2\right]^{-r}\ .
\ee
It is on this standard form, after the usual Wick rotation, that we impose the IR cutoff:
\be\label{cutoffint}
\left[\int {{\rm d}^dk\over (2\pi)^d} \
{\cal I}\right]_{{\rm IR-cutoff}\ \m} =  
i\, \int_{k_E^2\ge \m^2} {{\rm d}^dk_E\over (2\pi)^d}\,
Q(k_E-{\cal P}_E,p_s) \left[ k_E^2 + C -{\cal P}^2_E\right]^{-r}\ .
\ee
Note that the cutoff is applied on $k_E$ which is the Euclidean version of the {\it shifted} loop momentum $k'=k+{\cal P}(x_a,p_s)$. The shift, and hence the actual cutoff, depend on the Feynman parameters $x_a$ and the external $p_s$, but there is no arbitrariness and after integrating the $x_a$ we get a ``standard" result, i.e. independent of the arbitrariness in assigning momenta to the internal lines.\footnote{
There is another point one might worry about: when summing different diagrams that contribute to a given $\G_\m^{(n)}$ one might want to simplify expressions by combining terms involving different numbers of propagators before doing the loop-integral. Typically this occurs in theories that have
at the same time cubic and quartic vertices like non-abelian gauge theories (see e.g. eq. (\ref{quadraticA})). For example, one might want to rewrite ${1\over k^2+m^2}$ as ${(q-k)^2+m^2\over [k^2+m^2][(q-k)^2+m^2]}$. Applying our prescription for IR regularization to the integral of the second expression yields a complicated-looking integral over the Feynman parameter. One may nevertheless check explicitly that this exactly reduces to the IR regularized integral of the first expression. It is not completely clear to us whether such rewritings {\it always} yield the same result for the IR cutoff integrals in the end. To remove any ambiguity, the rule for a given diagram with $r$ propagators is to apply the IR cutoff directly on the corresponding expression with the
denominator $[{k'}^2 + ...]^{-r}$.
} 

It should be perfectly clear that one can use exactly the same prescription to impose a UV cutoff $\L$ on this standard form of any one-loop integral. Thus when using $\G_\m$ to compute correlation functions one just has to impose the UV cutoff $\m$ with this same prescription and is is obvious that one exactly obtains the part of the momentum integral that was left out when computing $\G_\m$.
Let us mention that instead of Feynman parameters one could have used parametric representation of the propagators, with a very similar result.

For later reference, let us note that the relevant one-loop integrals with the IR cutoff $\m$,
\be\label{mLintegraldim}
I_N(R)=i \int_{k_E^2\ge\m^2} {{\rm d}^d k_E\over (2\pi)^d}\ {1 \over (k_E^2 + R)^{N}} \qquad , \qquad
I_N^{\l\ldots\r}(R)=i \int_{k_E^2\ge\m^2} {{\rm d}^d k_E\over (2\pi)^d}\ {k_E^\l \ldots k_E^\r \over (k_E^2 + R)^{N}} \ ,
\ee
are given by (see appendix A.1)
\ba\label{mLint2dim}
I_1(R)
&=& {i\over (4\pi)^{d/2}\G({d\over 2})} 
\left( -{2\over \e}R -\m^2 + R \log(\m^2+R)+{\cal O}(\e)\right) \ ,
\nonumber\\
I_2(R)
&=& {i\over (4\pi)^{d/2}\G({d\over 2})} 
\left( {2\over \e} -{R\over \m^2+R} - \log(\m^2+R)+{\cal O}(\e)\right) \ ,
\nonumber\\
I_N(R)
&=& {i\over (4\pi)^{d/2}\G({d\over 2})} 
\Big({1\over N-2} {1\over (\m^2+R)^{N-2}}
-{1\over N-1} {R\over (\m^2+R)^{N-1}} +{\cal O}(\e) \Big) \ ,\quad N\ge 3 \ ,\quad \ \
\ea
where $\e=4-d$ and ${1\over (4\pi)^{d/2}\G({d\over 2})}={1\over (4\pi)^2}\left[ 1+\e\, {1-\g+\log 4\pi\over 2}+{\cal O}(\e^2)\right]$, 
as well as
\ba\label{tensorintdim}
I_N^{\l\r}(R)
&=& {1\over d}\ \delta^{\l\r} \left( I_{N-1}(R) - R I_N(R)\right) \ ,
\nonumber\\
I_N^{\n\l\r\s}(R)
&=& {1\over d(d+2)}\left(\delta^{\n\l}\delta^{\r\s} +\delta^{\n\r} \delta^{\l\s} +\delta^{\n\s}\delta^{\l\r}  \right)
\left( I_{N-2}(R) - 2R I_{N-1}(R) +R^2 I_N(R) \right)\ .\quad\ \ 
\ea
Of course, as already noted, the whole issue of UV-divergences and renormalization has to be addressed when computing the Wilsonian effective action $\G_\m$ just in the same way it had to be discussed when computing the 1PI effective action $\G$.

\subsection{Examples}

\subsubsection{Scalar $\vf^4$ theory in 4 dimensions}

As an explicit example, we apply this procedure to the one-loop diagram on the left of Fig. \ref{4pointfct}. Here we have two propagators, so that we just have a single Feynman parameter $x$. Furthermore $Q(k,P)=1$, ${\cal P}(x,P)=-x P$ and $C(x,P,m)=x P^2+m^2$. Then $R\equiv C-{\cal P}^2=m^2+x(1-x) P^2$. We are left with
\be
J_\m^{(4)}(P^2) 
=i\left[\int_0^1 \d x\ i\int {{\rm d}^dk_E\over (2\pi)^d} {\cal I}\Big\vert_{Fig. 1}\right]_{{\rm IR-cutoff}\ \m}  
=i\int_0^1 \d x \ I_2(m^2+x(1-x)P^2)
\ .
\ee
Inserting (\ref{mLint2dim}) and  $d=4-\e$ yields 
\be\label{J4m}
J_\m^{(4)}(P^2) = {1\over (4\pi)^2} \left\{
c+\int_0^1\d x\left[ \log(\m^2 +m^2+x(1-x) P^2) 
- { \m^2\over \m^2 + m^2+x(1-x) P^2} \right] + {\cal O}(\e)\right\}\ ,
\ee
where $c=-{2\over \e}+\g- \log 4\pi$.
Note that in the $\m\to 0$ limit, $J_\m^{(4)}(P^2)$ reproduces the standard one-loop contribution $J^{(4)}(P^2)$ to the 1PI four-point vertex, cf. (\ref{G4}), as it obviously should. 
To remove the ${2\over \e}$ pole and make $\G_\m^{(4)}$ finite at this order one has to add the ($\m$-independent) counterterm $\int {\rm d}^4 x\ \left(-{1\over 4!}\right) {3 g^2\over 2(4\pi)^2}\left({2\over \e}+c_0\right) \vf^4$, where the value of the finite constant $c_0$ depends on the renormalization condition. Then, up to order $g^2$ we have\be\label{GW41L}
\G^{(4)}_{\m, \rm tree}(p_i) + \G^{(4)}_{\m, \rm 1-loop}(p_i)
= -g\, - {g^2\over 2}
\left[J_\m^{(4)}((p_1+p_2)^2) + J_\m^{(4)}((p_1+p_3)^2) +
J_\m^{(4)}((p_2+p_3)^2) \right] \ ,
\ee
where $J_\m^{(4)}$ is still given by (\ref{J4m}) but now with $c=c_0+\g-\log4\pi$.
This example clearly shows several of the general features discussed above:
\begin{itemize}
\item
In a massless theory $\m$ is not just a fictitious mass: $J^{(4)}_\m$ for $m=0$ is {\it not} the same as $J^{(4)}$ with $m$ simply replaced by $\m$. 
\item
It is obvious from (\ref{J4m}) that the corresponding contribution to the Wilsonian effective action is indeed local, even for $m=0$. Explicitly, for $m=0$, one has $J_\m^{(4)}(P^2)\vert_{m=0}={1\over (4\pi)^2} \left\{ c +\log\m^2 + j^{(4)}\left({p^2\over \m^2}\right)\right\}$ with
\be\label{jfct}
j^{(4)}(z)=-2 +{2+z\over \sqrt{z(4+z)}} 
\log\left({1+\sqrt{z\over 4+z}\over 1-\sqrt{z\over 4+z}}\right)
= -1 +{z\over 3} -{z^2\over 20} +{\cal O}(z^3)
\ ,
\ee
which is free of singularities as long as $\vert z\vert <4$. 
\item
As discussed below eq. (\ref{cutoffint}), when computing correlation functions from the propagators and vertices given by the Wilsonian action $\G_\m$ (cf (\ref{GW41L})) one has to implement the {\it ultra-violet} cutoff $\m$ using exactly the same prescription. For the example of the 4-point vertex function at order $g^2$ one has to add the two contributions shown in Fig. \ref{4pointfct} (for each of the $s$, $t$ and $u$-``channels"). The tree-level contribution corresponding to the vertex $\sim g^2$ from $\G_{\m, \rm 1-loop}$ (right part of Fig. \ref{4pointfct}) is just $-{g^2\over 2} J^{(4)}_\m(P^2)$, while the contribution corresponding to the left part of Fig. \ref{4pointfct} is the loop-diagram, now with a UV cutoff $\m$,
involving two vertices $(-g)$ from $\G_{\m, \rm tree}$  . Hence, the latter contribution is finite and can be computed directly in $d=4$ giving $-{1\over 2}(-g)^2$ times
\ba\label{UVcutint}
i\int_0^1{\rm d}x\, i\int_{k_E^2\le \m^2} 
{{\rm d}^4 k_E \over (2\pi)^4} {1\over [k_E^2 + m^2 +x(1-x) P^2]^2}
=&&\hskip-6.mm
{1\over (4\pi)^2} \int_0^1{\rm d}x\,  
\Big[ \log {m^2+x(1-x)P^2\over \m^2+m^2+x(1-x)P^2}
\nonumber\\
&&\hskip0.8cm 
+ {\m^2 \over \m^2+m^2+x(1-x)P^2}\Big]\, .\ \ \ 
\ea
This contribution (\ref{UVcutint}) and $J^{(4)}_\m(P^2)$ as given in (\ref{J4m}) precisely add up to produce the order $g^2$ contribution $J^{(4)}(P^2)$ to the 1PI four-point vertex given in (\ref{G4}), as promised.
\end{itemize}

\noindent
Let us also note that, up to one loop, $\G_\m^{(2)}$ is given by
\be\label{G2oneloop}
\G^{(2)}_{\m, \rm tree} + \G^{(2)}_{\m, \rm 1-loop}
=-(p^2+m^2) + {g\over 2(4\pi)^2} \left(\m^2+ c_2\ m^2 -m^2\log(\m^2+m^2) \right)\ ,
\ee
where the value of the finite constant $c_2$ depends on the renormalization conditions. In minimal subtraction e.g. $c_2=1-\g+\log 4\pi$. Note that, even for $m=0$, there is a non-vanishing $\G^{(2)}_{\m, \rm 1-loop}={g\over 2(4\pi)^2}\m^2$. One can now compute the 1PI $\G^{(2)}$ up to order $g$, starting from the Wilsonian action $\G_\m=\G_\m^{(2)}\vert_{g^0}+\G_\m^{(2)}\vert_{g}+\G_\m^{(4)}\vert_{g}+\ldots$. It receives two contributions, a tree-level contribution with the 2-point vertex as given in (\ref{G2oneloop}) and a one-loop contribution with UV cutoff $\m$, involving the 4-point vertex $\G_\m^{(4)}\vert_{g}=-g$. The latter gives
\be\label{G2UVmu}
-{g\over 2} \int_{k_E^2\le\m^2} {{\rm d}^4k_E\over (2\pi)^4}
{1\over k_E^2+m^2} = -{g\over 2(4\pi)^2} \left(
\m^2-m^2\log{\m^2+m^2\over m^2}\right) \ .
\ee
When added to the former, the $\m$-dependence disappears and one reproduces the one-particle irreducible $\G^{(2)}=-(p^2+m^2)+{g \over 2(4\pi)^2} m^2\left( c_2-\log m^2\right)+{\cal O}(g^2)$.

In the appendix we compute the two-loop contributions to $\G_\m^{(2)}$. This will turn out to be quite a non-trivial example.

\subsubsection{Chiral fermions in 2 dimensions}

It is instructive to look at another example: consider a massless chiral fermion coupled to a $U(1)$ gauge field in {\it two} dimensions. At one loop,  its contribution to the vacuum-polarization $\G^{(2)}$ of the gauge field is given by the (anomalous) current two-point function 
$\langle j_+(p) j_+(-p)\rangle$. 
Its computation is straightforward and parallels e.g. the nice discussion of the two-point function of the energy-momentum tensor $T_{++}$ in ref. \cite{AGW}. Here we want to implement the IR-cutoff $\m$ on the momentum integral in the same way as we did above. The relevant  momentum integral then is
\be\label{jpjpint}
\int_{\m^2} {{\rm d}^d k\over (2\pi)^d}\ 
{k_+\over k^2}\ {p_+ +k_+\over (p+k)^2}
=i\int_0^1 {\rm d}x \int_{k_E^2\ge\m^2}\ {{\rm d}^d k_E\over (2\pi)^d} {k_E^+ k_E^+ -x(1-x) p_+^2\over [k_E^2+x(1-x)p^2]^2} \ ,
\ee
where $p_\pm=p_1\pm p_0$ (so that $p^2=p_+ p_-$) and $k_E^+=i k_{E,0}+k_{E,1}$. The integral of $k_E^+ k_E^+$ vanishes by symmetry and the remaining integral is convergent for $d=2$. One gets
\be\label{G2mu}
\G^{(2)}_\m\sim\langle j_+(p) j_+(-p)\rangle_\m 
=\hat c\ {p_+\over p_-}\ {\cal J}\left( {\m^2\over p^2}\right) \ ,
\ee
where
\be
{\cal J}(\xi) = \int_0^1 {\rm d}x\ {x(1-x)\over \xi+x(1-x)}\ .
\ee

Setting $\m=0$ gives back the corresponding part of the 1PI effective action. Since ${\cal J}(0)=1$ this is
\be\label{G2mu0}
\G^{(2)}\sim\langle j_+(p) j_+(-p)\rangle_{{}_{\m=0}} 
=\hat c\ {p_+\over p_-} \ .
\ee
This clearly shows the non-local character of the corresponding contribution to the 1PI action $\G$ and also
exhibits the usual anomaly $p_-\langle j_+(p) j_+(-p)\rangle =\hat c\ p_+ \ne 0$ (which in turn, as always, is local). 
For non-vanishing $\xi$, the integral ${\cal J}(\xi)$ is elementary (and similar to $j^{(4)}({1\over \xi})$ given above) and can be easily evaluated in the different regions $\xi<-1/4\ $,\ $\ -1/4<\xi<0$ and $\xi>0$. However, it is immediately obvious that for large $|\xi|$ it has a series expansion ${\cal J}(\xi) \sim {1\over 6\xi} + {\cal O}({1\over\xi^2})$. Thus for $\m^2\gg p^2\equiv p_+ p_-$ we get $\langle j_+(p) j_+(-p)\rangle_\m \sim \hat c\ {p_+\over p_-} {p^2\over 6\m^2}\left( 1+{\cal O}({p^2\over \m^2})\right) = \hat c \ {p_+^2\over 6\m^2}\left( 1+{\cal O}({p^2\over \m^2})\right)$ which clearly is local.
It is also obvious that, as a function of the real variable $\xi$, 
$\ \ {\cal J}(\xi)$ is everywhere decreasing since ${\cal J}'(\xi)<0$, and hence (since ${\cal J}(-\infty)={\cal J}(\infty)=0$) it must be singular somewhere. Indeed, ${\cal J}(\xi)$ is singular at $\xi=-{1\over 4}\ $: the expansion in inverse powers of $\m^2$ ceases to converge at $\m^2=-p^2/4$ and one could say that it is at this point where $\G^{(2)}_\m$ undergoes the transition from an infinite sum of local terms to a non-local expression.

\subsection{Symmetries of the Wilsonian action and (non)renomalization}

As we have discussed at length, to compute $\G_\m[\vf]$ we need to impose a UV regularization and to specify the IR cutoff $\m$. Just as for the 1PI effective action, any {\it linear} classical symmetry will be a symmetry also of the Wilsonian effective action if both, UV regularization and IR cutoff, preserve these symmetries. In particular, we have displayed cutoffs that preserve Lorentz invariance and, if present, supersymmetry. Consider first a non-gauge theory. The question of gauge invariance will be discussed below.
The Wilsonian effective action can be expanded in a series of terms with increasing numbers of derivatives, each of them being local and invariant under the non-anomalous symmetries. The $\m$-dependent coefficients of these terms are the Wilsonian coupling constants $g_n(\m)$. 
Accordingly, the Wilsonian $\b$-functions are defined as 
\be\label{betafct}
\b_n(g_m(\m))=\m {{\rm d}\over {\rm d}\m} g_n(\m) \ .
\ee
If one can show, using the symmetries of $\G_\m$, that certain couplings $g_n(\m)$ actually do not depend on $\m$ at all, then these couplings equal their bare values. This means that the corresponding proper vertices do not receive any contributions from loop diagrams (with infrared cutoff $\m$) or even non-perturbatively, i.e. they are not renormalized. This is typically the argument used in supersymmetric theories in \cite{SeibergNR} for the proof of the non-renormalization theorem for the $F$-terms. 

It is very important to realize that the Wilsonian couplings $g_n(\m)$
are different from the corresponding effective couplings $g_n(0)$ in the 1PI action and that the corresponding $\b$-functions\footnote{
Recall that the 1PI couplings $g_n(0)$ instead depend on the scale $\lambda$ used to define the renormalisation conditions and their $\b$-functions are defined as $\b_n=\l {{\rm d}\over {\rm d}\l} g_n(0)$.
} 
also are not the same. In theories involving massless fields, going from the Wilsonian couplings to the 1PI couplings one typically has to include terms that potentially receive infrared divergent contributions.
These questions have been extensively discussed in ref. \cite{SV}. 

\subsection{Gauge invariance of the Wilsonian effective action}

\subsubsection{Slavnov-Talor identities}

Obviously, if we are dealing with a gauge theory and if the gauge symmetry is not anomalous, the 1PI effective action must reflect the gauge invariance. As already discussed above, this is encoded in the Zinn-Justin equations which are a reflection of the BRST invariance of the gauge-fixed action. We also noted that if one uses a background field gauge the 1PI effective action really is gauge invariant. 

The gauge invariance of the Wilsonian effective action turns out to be a more complicated question. The basic point is that the introduction of the infrared scale $\m$ a priori breaks gauge, resp. BRST invariance. For example, it is well-known from the one-loop computations of the vacuum polarization in gauge theories that the introduction of an explicit momentum UV-cutoff generates quadratic divergences that lead to non-gauge invariant mass terms for the gauge fields. Clearly the same happens with an explicit infrared momentum cutoff $\m$. Alternatively, consider a BRST transformation like $s\, \psi \sim \eta \psi$. It is non-linear and hence is {\it not} diagonal in the momenta and the explicit introduction of the cutoff $\m$ is not manifestly BRST invariant. Thus one cannot automatically conclude that the Wilsonian effective action satisfies the Zinn-Justin equation, or equivalently that the appropriate Slavnov-Taylor identities are satisfied. 

The question of gauge invariance was much studied in the framework of  the exact renormalization group (ERG) \cite{Pol} using the flow equations. As already mentioned, in this context one computes with a UV cutoff $\L$ and deals with effective actions that have $\L$-dependent interactions. The basic point then is how to guarantee that the physical correlation functions obey the Ward identities and that the S-matrix is unitary. Probably the first gauge invariant UV regularization scheme involving an explicit scale $\L$ was constructed \footnote{
I am grateful to Bruce Campbell for bringing this reference to my attention.
} 
by Warr \cite{Warr} by adding ingeniously arranged higher covariant derivative terms $\sim \left( {D^2\over \L^2}\right)^n$ to the action. This allowed him to obtain regularized Ward identities for the regularized correlation functions which reduce to the standard Ward identities for the (finite) correlation functions in the limit $\L\to\infty$, thereby guaranteeing unitarity of the S-matrix. Although very interesting, this scheme is designed to study only physics at scales well below $\L$ where one effectively can consider the $\L\to\infty$ limit.
A somewhat different treatment was given by Bechi \cite{Bechi} who used a UV cutoff that breaks the gauge symmetry but showed that one can add appropriately fine-tuned non-invariant ($\L$-dependent) counterterms to the effective action in order to recover the Ward identities.
A more modern treatment following the same idea can be found in \cite{Sonoda}. These questions were also studied in detail in \cite{Kopper, Bonini} where it was shown that by exploiting the freedom in the choice of appropriate renormalization conditions,
the Ward identities are recovered at the end of the renormalization group flow. Said differently, the Ward identities receive $\L$-dependent corrections which flow to zero. Similarly, ref. \cite{Ellwanger} showed that the effective action with IR cutoff $\m$ obtained from the flow equations satisfies modified Slavnov-Taylor identities that reduce to the ordinary Slavnov-Taylor identities in the limit $\m\to 0$. Interesting as they are, these approaches only guarantee gauge invariance at the end point of the RG flow while we really would like to argue for gauge invariance at any finite scale $\m$.
More recently, refs. \cite{Rosten} have formulated  ERG flow equations for gauge theories in a manifestly gauge invariant way by realizing the cutoff $\L$ via a spontaneously broken larger gauge invariance. Finally,
we should mention that it is also possible to introduce explicit momentum cutoffs by using a lattice formulation even for chiral gauge theories without breaking gauge invariance \cite{Luscher} but, of course, the lattice breaks explicit Lorentz invariance.

As explained above, in this note we do not use the flow equations of the ERG and instead define the Wilsonian effective action $\G_\m$, starting from ordinary microscopic Yang Mills theory, by explicitly integrating out all the modes above the scale $\m$. We want to see whether in some cases this could still lead to a gauge invariant $\G_\m$ for any finite $\m$.

\subsubsection{Background field gauge}

Again, in order to be able to argue for gauge invariance of $\G_\m$, it is more convenient to work in background field gauge. This is the procedure adopted throughout refs. \cite{SV} for their study of supersymmetric gauge theories. As far as the UV regularization is concerned, these references use a combination of Pauli-Villars for the chiral multiplets and higher-derivative regularization for the vector multiplets. They do not, however, explicitly specify the way they implement the IR-cutoff $\m$. Note also that the above-mentioned gauge invariant regularization by Warr has been extended to background field gauge in ref. \cite{KL}. Here, we will use the explicit IR-cutoff $\m$ introduced above which has the advantage of having a clear and intuitive interpretation, and which can take any finite value: $\G_\m$ has a well-defined meaning whether the external momenta satisfy $p^2\lesssim \m^2$ or not. As already mentioned, this IR cutoff explicitly breaks gauge invariance by generating e.g. mass terms $\sim \m^2$ for the gauge fields. In the next section, we will proceed to an explicit one-loop computation of several terms in the Wilsonian effective action $\G_\m$ for general gauge theories. We will see that not only these mass terms are indeed present, but actually there are (infinitely) many other non-gauge invariant terms in the effective action for a generic gauge theory.\footnote{
Let us insist that, although in a general gauge theory $\G_{\m, \rm 1-loop}$ is not gauge invariant, the one-loop correlation functions computed from this $\G_{\m, \rm 1-loop}$ are the $\m$-independent 1PI correlation functions that do satisfy the Ward identities.}
We will give a complete one-loop computation of these terms that are bilinear in the gauge fields and involve arbitrarily many derivatives. However, we will also show that in a {\it supersymmetric} theory all these non-gauge-invariant terms in $\G_\m$ {\it cancel} within each supermultiplet. We take this as strong evidence that the same cancellation of the non-gauge invariant terms due to supersymmetry  occurs for the full Wilsonian effective action which then is indeed {\it Lorentz, susy and gauge-invariant} for all $\m$, and  can be expanded, as long as $p^2\lesssim \m^2$,  as an (infinite) sum of {\it local} terms.

\subsubsection{Anomalies}

One more point we should discuss here concerns possible anomalies. In general we will be interested in theories that contain chiral fermions, potentially leading to gauge or global anomalies. A simple explicit example was discussed in section 2.4.2. Of course, gauge anomalies render the theory inconsistent and (as usual) we will suppose that the matter content is arranged in such a way that they cancel. However, anomalies in global symmetries often play an important role. An anomaly is a non-invariance of the effective action that cannot be removed by adding local counterterms to the classical action.  As we have seen in the two-dimensional example above, the non-invariant terms in $\G$ must be non-local since if they were local one could just subtract these terms from the classical action as local counterterms, and the new effective action would be invariant. At first sight it then seems as if no anomaly could manifest itself in the Wilsonian effective action (at least for large enough $\m$) , and that it is only produced as an IR effect when going from $\G_\m$ to $\G$. This is not true, however, since although $\G_\m$ is a sum of local terms, even the non-invariant part a priori is an {\it infinite} sum, and so one would have to add infinitely many counterterms to the classical action. More important, these counterterms all have coefficients that depend on $\m$. For a fixed value of $\m$ they would lead to an invariant new $\tilde\G_\m$, but if we compute $\tilde\G_{\m'}$ at another scale $\m'\ne\m$ the non-invariant terms would no longer cancel. Thus there is no way to cancel the anomaly in $\G_\m$ for arbitrary $\m$ by adding ($\m$-independent) local counterterms to the classical action, and it makes perfectly sense to discuss global (or gauge) anomalies at the level of the Wilsonian effective action.

\section{One-loop Wilsonian action for gauge theories, non-gauge invariant terms and their cancellation in susy theories}
\setcounter{equation}{0}

We will now explicitly compute, at one loop, various terms of the Wilsonian effective action for general gauge theories. As discussed above, we will do this in background field gauge.
It is by now a standard textbook computation using background field gauge to obtain the coefficient of the $F_{\mu\nu} F^{\mu\nu}$ term in the 1PI effective action of gauge theories coupled to spin ${1\over 2}$ Dirac  fields transforming in some representation $R$ of the gauge group, thereby deriving the celebrated $\b$-function. Here we will follow the presentation and computation of \cite{Weinbook}, and adapt it by introducing the explicit IR cutoff $\m$ according to our prescription explained in section 2.3. We will first compute terms quadratic and quartic in a constant background gauge field and then quadratic terms in an arbitrary background gauge field. In all cases we will find many terms that are not gauge invariant. However, we will also see that, in supersymmetric gauge theories, these non-invariant terms cancel when adding the contributions of all fields in any supermultiplet. Using our results, we will give explicit formulae for the one-loop Wilsonian couplings for all higher-derivative terms $\sim F D^{2n} F$ in the Wilsonian effective action in arbitrary supersymmetric gauge theories.

\subsection{Quadratic and quartic terms for constant background gauge fields}

We will first compute the one-loop  Wilsonian effective action up to quartic order in the (background) gauge fields at zero momentum, i.e. for constant fields $A$, and at vanishing ghost and fermion field background. Then
$F_{\m\n}=-i[A_\m,A_\n]$ and
$\tr F_{\m\n}F^{\m\n} = 2\tr (A_\m A^\m A_\n A^\n 
- A_\m A_\n A^\m A^\n)$.
After going through the background gauge fixing procedure, the one-loop effective action is given by the logarithm of the product of determinants of the propagators, in the presence of the background fields, of the gauge  ($A'$), ghost ($\o'$) and fermionic matter ($\p'$) fields.\footnote{
In susy gauge theories one also has scalars. Their contributions are similar to those of the ghost fields and will be given later on.
}  
Since the latter are taken to be constant, the determinants are easily evaluated. Explicitly one has
\ba\label{Goneloop}
\G_{\rm 1-loop}[A]
&=& \int {\rm d}^4 x \
\g_{\rm 1-loop}[A] \ ,
\nonumber\\
i\g_{\rm 1-loop}[A]
&=&\int {{\rm d}^4 p\over (2\pi)^4}
\left[ -{1\over 2}\tr\log\cM^{A'}(p) +\tr\log\cM^{\o'}(p) + \tr\log\cM^{\p'}(p) \right] \ ,
\ea
where in Feynman gauge ($\xi=1$) \cite{Weinbook}
\ba\label{Mop}
\cM^{A'}_{\m\n}(p) 
&=& \eta_{\m\n} p^2 -2\eta_{\m\n}p_\l A^\l +\eta_{\m\n} A_\l A^\l 
+2i F_{\m\n} \ ,
\nonumber\\
\cM^{\o'}(p) &=& p^2 -2p_\l A^\l  + A_\l A^\l \ ,
\nonumber\\
\cM^{\p'}(p)
&=& i \psl + m -i \Asl
\ .
\ea
Note that the traces in (\ref{Goneloop}) are traces over Lorentz indices, Dirac matrices and Lie algebra generators. As usual, $A_\n=A_\n^a t_a$ with the generators $t_a$ in the adjoint for the gauge and ghost fields and in some matter representation $R$ for the fermions. 
To compute the logarithms in (\ref{Goneloop}), we split each $\cM$
as $\cM=\cM_0 +\cM_1 +\cM_2$ where $\cM_0^{-1}$ is the free propagator and $\cM_1$, resp. $\cM_2$ are linear, resp. bilinear in the background gauge field $A$. Using the formula $\tr\log\cM=\tr\log\cM_0 -\sum_{n=1}^\infty {(-)^n\over n} \tr \left( \cM_0^{-1} (\cM_1+\cM_2)\right)^n$
it is easy to pick out the contributions to the terms in the effective action involving a given number\footnote{
The term without any field dependence in $i\g_{\rm 1-loop}$ is just
given by replacing the full $\cM$'s in (\ref{Goneloop})  by the $\cM_0$'s which gives 
$\int {{\rm d}^4 p\over (2\pi)^4} \left[(-{1\over 2} 4+1) (-\log p^2) \tra 1 -2\log(p^2+m^2) \trR 1 \right]$, with the $2$ in front of the second $\log$ absent for Majorana fermions. Thus we see that the integrand vanishes if the fermions are massless Majorana fermions and transform also in the adjoint representation as is the case for the ${\cal N}=1$ vector multiplet: the vacuum energy density vanishes as it should for unbroken susy. This result is not affected by the introduction of the IR cutoff $\m$.
} 
of gauge fields $A$.

\subsubsection{Quadratic terms}

{\it  General gauge theories}
\vskip 2.mm

\noindent
First, any  terms involving odd powers of $A$ obviously will vanish by Lorentz invariance (as we indeed use a Lorentz invariant UV and IR regularization) since at zero momentum there is no way to form a Lorentz scalar with an odd number of gauge fields $A$. Next, we look at the term quadratic in $A$. If present at zero momentum such a term clearly represents a mass term for the gauge field and breaks gauge invariance. The corresponding contribution to $\g_{\rm 1-loop}$ is
\ba\label{quadraticA}
i\g_{\rm 1-loop}\Big\vert_{A^2}
&=& \int {{\rm d}^4 p \over (2\pi)^4}\Big\{
\left(-{1\over 2} \eta_{\m\n}\eta^{\m\n} +1\right)
\left( {\eta^{\l\r}\over p^2} - {2 p^\l p^\r\over p^4}\right) 
\tra A_\l A_\r 
\nonumber\\
&&\hskip1.5cm + 2 \left( {\eta^{\l\r}\over p^2+m^2} - {2 p^\l p^\r\over (p^2+m^2)^2}\right) 
\trR A_\l A_\r  \Big\} \ .
\ea
Here the first line contains the contributions from $\cM^{A'}$ ($\sim 
-{1\over 2} \eta_{\m\n}\eta^{\m\n}$) and $\cM^{\o'}$  ($\sim 1$),
while the second line contains those of $\cM^\p$. Note that the factor of $2$ in the second line would be absent for Majorana fermions.
In dimensional regularization {\it without} any IR cutoff $\m$ one has 
\be\label{a2dim0}
\int {\rm d}^d p \left( {\eta^{\l\r}\over p^2+m^2} - {2 p^\l p^\r\over (p^2+m^2)^2}\right) = 0 \ ,
\ee
for all $m$ and in particular also for $m=0$. This implies the vanishing of (\ref{quadraticA}) and the absence of mass terms for the gauge field in the 1PI action $\G$.

In section 2.3, we discussed how to introduce the IR cutoff $\m$ on the standard form of the loop integrals. To bring them into standard form one had to perform translations of the integration variables and in order to be able to do so we had to work with already convergent integrals. This is why we used dimensional regularization of the integrals. Here, however, since we work at vanishing external momentum, the integrals already are in standard form. Thus, alternatively, we can simply introduce a 
euclidean momentum UV cutoff $\L$ and IR cutoff $\m$ on the integrals in (\ref{quadraticA}), working directly in 4 dimensions. It will be interesting to compare both UV regularizations. The relevant integrals (which we denote by $\hat I_N$ to distinguish them from their dimensionally regularized cousins $I_N$) then are 
\be\label{mLintegral}
\hat I_N(m^2)=i \int_{\m^2\le p_E^2\le\L^2} {{\rm d}^4 p_E\over (2\pi)^4}\ {1\over (p_E^2 + m^2)^{N}}
\quad , \quad
\hat I_N^{\l\ldots\r}(m^2)=i \int_{\m^2\le p_E^2\le\L^2} {{\rm d}^4 p_E\over (2\pi)^4}\ {p_E^\l \ldots p_E^\r \over (p_E^2 + m^2)^{N}}\ ,
\ee
and are given by
\ba\label{mLint2}
\hat I_1(m^2)
&=& {i\over (4\pi)^2} \left( \L^2-\m^2 + m^2 \log{m^2+\m^2\over m^2 +\L^2}\right) \ ,
\nonumber\\
\hat I_2(m^2)
&=& {i\over (4\pi)^2} \left( {m^2\over \L^2+m^2} - {m^2\over \m^2+m^2} - \log{m^2+\m^2\over m^2 +\L^2}\right) \ ,
\nonumber\\
\hat I_N(m^2)
&=& {i\over (4\pi)^2} \Big(  {1\over N-2} {1\over (\m^2+m^2)^{N-2}}
-{1\over N-1} {m^2\over (\m^2+m^2)^{N-1}} +{\cal O}({1\over \L}) \Big) \ ,\quad N\ge 3 \ ,\quad
\ea
as well as
\ba\label{tensorint}
\hat I_N^{\l\r}(m^2)
&=& {1\over 4}\delta^{\l\r} \left( \hat I_{N-1}(m) - m^2 \hat I_N(m)\right) \ ,
\nonumber\\
\hat I_N^{\n\l\r\s}(m^2)
&=& {1\over 24}\left(\delta^{\n\l}\delta^{\r\s} +\delta^{\n\r} \delta^{\l\s} +\delta^{\n\s}\delta^{\l\r}  \right)
\left( \hat I_{N-2}(m) - 2m^2\hat  I_{N-1}(m) +m^4 \hat I_N(m) \right)\ .\ \ \ \
\ea
Using these integrals we get from (\ref{quadraticA})
\ba\label{quadraticA2}
i\g_{\m,\rm 1-loop}\Big\vert_{A^2}
&=& \Big\{
\left(-2 +1\right)
 {1\over 2} \hat I_1(0)\tra A^\l A_\l 
\nonumber\\
&&
\ +\  \left( \hat I_1(m^2)  + m^2 \hat I_2(m^2))\right) 
\trR A^{\l} A_\l  \Big\} \ ,
\ea
with the first line coming from the gauge field and ghost loop and the second line from the fermion matter loop. Explicitly one has
\be\label{quadraticA3}
i\g_{\m,\rm 1-loop}\Big\vert_{A^2}
={i\over (4\pi)^2} 
\left\{-{1\over 2} (\L^2-\m^2) \tra A^\l A_\l
+  \left( \L^2-{\m^4\over \m^2+m^2}-m^2\right) 
\trR A^{\l} A_\l  +{\cal O}({1\over \L})\right\} \ .
\ee
Had we used dimensional regularization together with the IR cutoff $\m$, the quadratic UV divergences would have been absent and we would have obtained instead
\be\label{quadraticA4}
i\g_{\m,\rm 1-loop}\Big\vert_{A^2}
={i\over (4\pi)^2} 
\left\{{\m^2\over 2} \tra A^\l A_\l
-{\m^4\over \m^2+m^2}
\trR A^{\l} A_\l   +{\cal O}(\e)\right\} \ .
\ee
Both expressions (\ref{quadraticA3}) and (\ref{quadraticA4}) in general are non-vanishing for $\m\ne 0$, thus explicitly breaking the gauge invariance of the Wilsonian effective action.\footnote{
Note that (\ref{quadraticA3}) only has quadratic divergences, while the logarithmic divergences have cancelled. Similarly (\ref{quadraticA4}) has no divergences at all. Nevertheless, the finite part of (\ref{quadraticA3}) does not equal (\ref{quadraticA4}), since the latter gets an extra finite contribution $\sim (4-d) {2\over \e} m^2$, which can be traced to the difference between (\ref{tensorintdim}) and (\ref{tensorint}). Note also that we mentioned in sect. 2.3 that one can opt for dimensional reduction (i.e. first doing all $\g$-matrix and tensor algebra in four dimensions and only then dimensionally continuing the integral) or ordinary dimensional regularization. In the latter case, in (\ref{quadraticA}) one should set $\eta_{\m\n}\eta^{\m\n}=d$ and the $2$ in the second line becomes $2^{1-\e/2}$. However, this only changes the ${\cal O}(\e)$ term in (\ref{quadraticA4}).
} 
Note that these non-invariant terms get contributions not only from the massless gauge and ghost fields but also from the massive fermion fields. This is somewhat contrary to the naive expectation that for a massive field it does not matter whether one imposes an IR cutoff or not. What remains true is that in dimensional regularization for $m\gg\m$ the non-invariant contribution of the massive fields is suppressed by a factor ${\m^2\over m^2}$.

\vskip2.mm
\noindent
{\it Supersymmetric gauge theories}
\vskip2.mm

\noindent
Now consider what happens in a supersymmetric gauge theory. We already noted in the footnote above that the vacuum energy density vanishes, independently of the IR-cutoff, thus leaving supersymmetry unbroken.
Concerning the quadratic terms, first look at the vector multiplet. The contributions of the gauge fields and the ghosts are unaltered, but for the fermions there are several modifications: they are massless Majorana fermions in the adjoint representation. Thus in the second line in (\ref{quadraticA2})
we must set $m=0$, include an extra factor ${1\over 2}$ and replace
$\trR A^{\l} A_\l$ by $\tra A^\l A_\l$. As a result $i\g_{\m,\rm 1-loop}\vert_{A^2}$ vanishes. Alternatively, this can be seen directly from (\ref{quadraticA}). Similarly for a chiral multiplet we have a complex boson and a Majorana fermion of the same mass $m$ and in the same representation $R$. The contribution of the complex boson can be obtained from that of the ghost in (\ref{quadraticA}) by replacing the massless propagator by a massive one, replacing  $\tra A^\l A_\l$ by  $\trR A^{\l} A_\l$ and changing the overall sign. Including a factor ${1\over 2}$ for the Majorana fermions it is then immediately clear from (\ref{quadraticA}) that both contributions cancel. Thus 
\be\label{quadraticsusy}
\g_{\m,\rm 1-loop}\Big\vert_{A^2}^{\rm vector\ multiplet}
=\g_{\m,\rm 1-loop}\Big\vert_{A^2}^{\rm chiral\ multiplet} = 0 \quad , 
\quad {\rm even\ for\ } \m\ne 0 \ .
\ee
This is valid whether the UV divergences have been regularized dimensionally or by the explicit cutoff $\L$.

\subsubsection{Quartic terms}

{\it  General gauge theories}
\vskip2.mm

\noindent
Next, we look at the terms in $\g_{\m,\rm 1-loop}$ that are quartic in the background gauge field. In particular this will yield the $\m$-dependence of the one-loop Wilsonian gauge coupling. We will also find non gauge-invariant terms in the presence of massive matter, that cancel however when considering full susy multiplets. Again, we first use an explicit UV cutoff $\L$ and an explicit IR cutoff $\m$ and work in 4 dimensions, and below compare with dimensional regularization. After the relevant algebra one finds for the gauge and ghost contributions
\ba\label{quarticA'}
i\g_{\m,\rm 1-loop}\Big\vert_{A^4}^{A'}
&=&-{1\over 2}\tra\Big\{ -2 \left((A_\l A^\l)^2 +F_{\l\r}F^{\l\r}\right) \hat I_2(0)+16 A_\n A_\r A_\l A^\l \hat I_3^{\n\r}(0) 
\nonumber\\
&&\hskip1.5cm- 16 A_\n A_\r A_\l A_\s \hat I_4^{\n\r\l\s}(0) \Big\}
\nonumber\\
&=&{5\over 6}\ \hat I_2(0)\, \tra F_{\l\r}F^{\l\r} \ ,
\ea
and
\ba\label{quartico'}
i\g_{\m,\rm 1-loop}\Big\vert_{A^4}^{\o'}
&=&\tra\Big\{ -{1\over 2} (A_\l A^\l)^2 \hat I_2(0)+ 4 A_\n A_\r A_\l A^\l \hat I_3^{\n\r}(0) 
- 4 A_\n A_\r A_\l A_\s \hat I_4^{\n\r\l\s}(0) \Big\}
\nonumber\\
&=&{1\over 12}\ \hat I_2(0)\, \tra F_{\l\r}F^{\l\r} \ ,
\ea
which are both manifestly gauge invariant, even though the $\hat I_N(0)$ are the IR and UV cutoff integrals (\ref{mLint2}). Next we give the contribution of the Dirac fermion of mass $m$. After some standard but lengthy algebra one finds
\ba\label{quarticp'}
i\g_{\m,\rm 1-loop}\Big\vert_{A^4}^{\p'}
&=&
-{1\over 4}\int_\m^\L {{\rm d}^4p\over (2\pi)^4} {\tr (\psl+im)\g^\n(\psl+im)\g^\r(\psl+im)\g^\l(\psl+im)\g^\s\over (p^2+m^2)^4} \trR A_\n A_\r A_\l A_\s
\nonumber\\
&=&
{1\over 3}\left( 2\hat I_2(m^2)+ 2m^2\hat I_3(m^2)-m^4\hat I_4(m^2)\right) \trR A_\l A_\r A^{\l}A^{\r}
\nonumber\\
&&\hskip-3.mm -{1\over 3}\left( 2\hat I_2(m^2)+ 2m^2\hat I_3(m^2)+2m^4\hat I_4(m^2)\right) \trR A_\l A^{\l} A_\r A^{\r}
\nonumber\\
&=& -{1\over 3}\left( \hat I_2(m^2)+ m^2\hat I_3(m^2)+m^4\hat I_4(m^2)\right) 
\trR F_{\r\l} F^{\r\l}
-m^4 \hat I_4(m^2) \trR A_\l A_\r A^{\l}A^{\r}
\ .
\nonumber\\
&& 
\ea
\vskip-2.mm
\noindent
In the massless case one just gets the gauge invariant $-{1\over 3}\, \hat I_2(0) \trR F_{\r\l} F^{\r\l}$ reproducing together with (\ref{quarticA'}) and (\ref{quartico'}) the well-known $\b$-function. However, for $m\ne 0$ we also get a non-gauge invariant term.\footnote{
Note that when using dimensional regularization {\it without} IR cutoff there is a subtle cancellation and this non-invariant term is absent.
Indeed, in dimensional regularization, in the second expression (\ref{quarticp'}) the coefficient of $\trR A_\l A_\r A^{\l}A^{\r}$ is replaced by ${1\over 3}\left( (2-{5\over 12}\e)I_2(m^2)+ 2m^2I_3(m^2)-m^4I_4(m^2)\right)$ and that of $\trR A_\l A^{\l} A_\r A^{\r}$
is replaced by $-{1\over 3}\left( (2-{2\over 3}\e)I_2(m^2)+ 2m^2I_3(m^2)+2m^4I_4(m^2)\right)$, where the $I$ now are the corresponding dimensionally regularized integrals. The ${\cal O}(\e)$ terms in front of $I_2$ together with the ${1\over \e}$ pole of $I_2$ now produce another non-gauge-invariant term $\sim {1\over 3} ({5\over 12}\e-{2\over 3}\e) {i\over (4\pi)^2} {2\over \e} = -{i\over 6(4\pi)^2}$ which exactly cancels the non-invariant term $\sim m^4 I_4(m^2)= {i\over 6(4\pi)^2}$.
}
We will also need the contribution of the complex scalar $\phi$. Again, this is similar to the contribution of the ghost, but with non-zero mass, an extra minus sign and the replacement $\tra \to \trR$. We obtain
\be\label{quarticcomplscalar}
i\g_{\m,\rm 1-loop}\Big\vert_{A^4}^{\phi'}
= -{1\over 12}\left( \hat I_2(m^2) -2m^2 \hat I_3(m^2) -2m^4 \hat I_4(m^2)\right) 
\trR F_{\l\r}F^{\l\r}
+{m^4\over 2} \hat I_4(m^2) \trR A_\l A_\r A^\l A^\r \ ,
\ee
again with a non gauge-invariant term $\sim m^4 \hat I_4(m^2)$. 

Thus again, just as what happened for the quadratic terms, the quartic terms in the Wilsonian effective action are not gauge-invariant due to the introduction of the explicit IR-cutoff. Somewhat contrary to naive expectations, the contributions to the non-invariant quartic terms only come from massive fields and not from the massless ones.

\vskip2.mm
\noindent
{\it Supersymmetric gauge theories}
\vskip2.mm

\noindent
How does this conclusion get modified in a supersymmetric theory? 
For the vector multiplet, the above computation applies but with the fermion mass set to zero, taking $R$ to be the adjoint representation, and including a factor ${1\over 2}$ for the Majorana fermions. Since $m=0$, all terms are gauge invariant. Similarly, for the chiral multiplet, the non-invariant terms cancel between the Majorana fermion and the complex scalar. We get
\be\label{quartivectormult}
i\g_{\m,\rm 1-loop}\Big\vert_{A^4}^{\rm vector\ multiplet}
=\left({5\over 6}+{1\over 12} -{1\over 6}\right) \hat I_2(0) \tra F_{\l\r}F^{\l\r} 
={3\over 4}\ \hat I_2(0) \tra F_{\l\r}F^{\l\r}\ ,
\ee
and
\be\label{quartichiralmult}
i\g_{\m,\rm 1-loop}\Big\vert_{A^4}^{\rm chiral\ multiplet}
=-{1\over 4}\ \hat I_2(m^2) \trR F_{\l\r}F^{\l\r}\ ,
\ee
all obviously again gauge invariant.

For gauge theories with extended supersymmetry, note that the ${\cal N}=2$ vector multiplet consists of an ${\cal N}=1$ vector and an ${\cal N}=1$ chiral multiplet (all in the adjoint), while the ${\cal N}=2$ hyper multiplet consists of two ${\cal N}=1$ chiral multiplets in the same representation $R$. Hence
\be\label{N=2}
i\g_{\m,\rm 1-loop}\Big\vert_{A^4}^{{\cal N}=2\ \rm vector}
={1\over 2}\ \hat I_2(0) \tra F_{\l\r}F^{\l\r}\quad , \quad
i\g_{\m,\rm 1-loop}\Big\vert_{A^4}^{{\cal N}=2\ \rm hyper}
=-{1\over 2}\ \hat I_2(m^2) \trR F_{\l\r}F^{\l\r} \ .
\ee
Of course, the ${\cal N}=4$ multiplet consists of an ${\cal N}=2$ vector and a massless hyper multiplet in the adjoint, and hence
\be\label{N=4}
i\g_{\m,\rm 1-loop}\Big\vert_{A^4}^{{\cal N}=4} = 0 \ .
\ee 

\vskip5.mm
\noindent
{\it Using dimensional regularization instead}
\vskip2.mm

\noindent
It is again interesting to see how these conclusions are modified if instead of the UV cutoff $\L$ we use dimensionally regularized integrals. We can treat the usual dimensional regularization (with all tensor and $\g$-matrix algebra in $d$ dimensions) and dimensional reduction (with $\g$-matrix and tensor algebra in 4 dimensions and only the integrals dimensionally regularized) simultaneously: the only difference is a factor $d_A=\eta_{\n\s}\eta^{\n\s}$ appearing in the $A'$ contribution ($d_A=d$, resp. 4), and a factor $d_\p$ in the $\p'$ contribution from the trace over $\g$-matrices ($d_\p=2^{d/2}$, resp. 4). One then finds (the relevant integrals are now given by (\ref{mLint2dim}) and (\ref{tensorintdim}))
\ba\label{quarticA'dim}
i\g_{\m,\rm 1-loop}\Big\vert_{A^4}^{A'}
&=&\left[1-{d_A\over 24}\left( 1+{2\e\over 3}\right)\right] I_2(0)\,
\tra F_{\l\r}F^{\l\r}
- {d_A\over 24} {i\over (4\pi)^2} \tra A_\l A_\r A^\l A^\r \ ,
\\
\nonumber\\
\label{quartico'dim}
i\g_{\m,\rm 1-loop}\Big\vert_{A^4}^{\o'}
&=&
\left({1\over 12}+{\e\over 18}\right) I_2(0)\, \tra F_{\l\r}F^{\l\r}
+{i\over 12 (4\pi)^2} \tra A_\l A_\r A^\l A^\r \ ,
\\
\nonumber\\
\label{quarticp'dim}
i\g_{\m,\rm 1-loop}\Big\vert_{A^4}^{\p'}
&=&
{d_\p\over 2}\left[ -\left({1\over 6}-{\e\over 18}\right) I_2(m^2)
-{m^2\over 6} I_3(m^2) -{m^2\over 6} I_4(m^2)\right]\, 
\trR F_{\l\r}F^{\l\r}
\nonumber\\
&&\hskip-2.mm
+{d_\p\over 2}\left[{i\over 12 (4\pi)^2}-{m^2\over 2} I_4(m^2)\right] \trR A_\l A_\r A^\l A^\r \ ,
\\
\nonumber\\
\label{quarticphi'dim}
i\g_{\m,\rm 1-loop}\Big\vert_{A^4}^{\phi'}
&=&
\left[ -\left({1\over 12}+{\e\over 18}\right) I_2(m^2)
+{m^2\over 6} I_3(m^2) +{m^2\over 6} I_4(m^2)\right]\, 
\trR F_{\l\r}F^{\l\r}
\nonumber\\
&&\hskip-2.mm
-\left[{i\over 12 (4\pi)^2}-{m^2\over 2} I_4(m^2)\right] \trR A_\l A_\r A^\l A^\r \ .
\ea
This time, each contribution contains non-gauge invariant terms.\footnote{ 
It is amusing to remark that there are more non-gauge invariant terms when using dimensional regularization than in the computation above done with an explicit UV cutoff.
}
Again, in a supersymmetric theory, they exactly cancel for the ${\cal N}=1$ vector and chiral multiplets, provided $d_A=d_\p=4\ $: As expected, we must use the dimensional reduction procedure rather than the usual dimensional regularization. We then get:
\ba\label{quarticvectdim}
i\g_{\m,\rm 1-loop}\Big\vert_{A^4}^{{\cal N}=1\ \rm vector}
&=&\hskip3.mm {3\over 4}\ I_2(0) \tra F_{\l\r}F^{\l\r}\ ,
\\
\label{quarticchiraldim}
i\g_{\m,\rm 1-loop}\Big\vert_{A^4}^{{\cal N}=1\ \rm chiral}
&=&-{1\over 4}\ I_2(m^2) \trR F_{\l\r}F^{\l\r}\ .
\ea
Not only are these gauge invariant, they also take exactly the same form as the corresponding quantities derived above with the UV cutoff $\L$ (i.e. all the extra terms $\sim \e I_2$ present in the individual non-supersymmetric contributions have also cancelled).

\subsection{Quadratic terms for non-constant background gauge fields}

\subsubsection{General gauge theories}

One might wonder whether the cancellation of the non-gauge invariant terms within the supermultiplets is a special feature for constant background fields or whether it continues to hold even at non-zero momenta. We will now compute the one-loop Wilsonian effective action $\G_{\m,\, \rm 1-loop}$ in the presence of the IR-cutoff $\m$ up to second order in an arbitrary {\it non}-constant gauge field $A_\n(x)$, thus involving arbitrarily many derivatives. It is not difficult to adapt the previous computation to the present case, and we will skip most of the details. Again, $\G_{\m,\,\rm 1-loop}$ is given by the sum of logarithms of the various determinants, but now all the momentum modes\footnote{
Our normalisation is $A_\n(x)
=\int{ {\rm d}^4 q\over (2\pi)^4}\ e^{-iq_\l x^\l} \tilde A_\n(q)$ so that the tree-level action 
$S_{\rm YM}=-{1\over 4g^2}\int {\rm d}^4 x\, F^a_{\l\s}F_a^{\l\s}$ 
corresponds to $\G^{(2)}_{\l\n}(q)\vert_{\rm tree}={1\over g^2}(q_\l q_\n - q^2 \eta_{\l\n}){1\over C_{\rm adj}}$, where $\tra t^a t^b=C_{\rm adj} \delta^{ab}$. Of course, we want to consider background gauge fields that are non-vanishing only for ${q^2\over \m^2}\lesssim 1$ in order to get a Wilsonian action that can be expanded in powers of ${q^2\over \m^2}$.
} 
$\tilde A_\n(q)$ will contribute. We write:
\be\label{Goneloopq}
\G_{\m,\,\rm 1-loop}\Big\vert_{A^2}= {1\over 2}
\int{ {\rm d}^4 q\over (2\pi)^4}\ \G^{(2)}_{\l\n}(q)\
\tr \tilde A^\l(q) \tilde A^\n(-q)  \ ,
\ee
where it is implicitly understood that $\G^{(2)}_{\l\n}(q)$ is meant to be one loop and that the trace is to be taken in the appropriate representation for each contribution. Explicitly, the different fields contribute
\ba\label{diffa}
\hskip-1.5cm
{i\over 2}\,\G^{(2)}_{\l\n}(q)\Big\vert^{A'}
&=&-{1\over 2}\Big\{\eta_{\r\s}\eta^{\r\s} \int{ {\rm d}^d p\over (2\pi)^d} 
\left({\eta_{\l\n}\over p^2} -{1\over 2} {(2p+q)_\l (2p+q)_\n\over p^2 (p+q)^2} \right) 
\nonumber\\
&&
\hskip8.mm +4\, (q_\l q_\n-q^2\eta_{\l\n}) \int{ {\rm d}^d p\over (2\pi)^d} {1\over p^2 (p+q)^2} \Big\} \ ,
\\
\nonumber\\
\label{diffgh}
\hskip-1.5cm
{i\over 2}\,\G^{(2)}_{\l\n}(q)\Big\vert^{\o'}
&=&\int{ {\rm d}^d p\over (2\pi)^d} 
\left({\eta_{\l\n}\over p^2} -{1\over 2} {(2p+q)_\l (2p+q)_\n\over p^2 (p+q)^2} \right) \ ,
\ea
\ba\label{diffpsi}
{i\over 2}\,\G^{(2)}_{\l\n}(q)\Big\vert^{\p'}
&=&
2\, \int{ {\rm d}^d p\over (2\pi)^d} {-p_\l (p+q)_\n - p_\n(p+q)_\l 
+p\cdot(p+q)\eta_{\l\n} +m^2\eta_{\l\n}\over [p^2+m^2][(p+q)^2+m^2]} \ ,
\\
\nonumber\\
\label{diffphi}
{i\over 2}\,\G^{(2)}_{\l\n}(q)\Big\vert^{\phi'}
&=&
- \int{ {\rm d}^d p\over (2\pi)^d} 
\left({\eta_{\l\n}\over p^2+m^2} -{1\over 2} {(2p+q)_\l (2p+q)_\n\over [p^2+m^2][(p+q)^2+m^2]} \right)\ ,
\ea
where, as before, $\phi'$ is a complex scalar and $\p'$ is a Dirac fermion. For Majorana fermions the factor $2$ in (\ref{diffpsi}) is absent. 

This time we have opted for dimensional regularization (actually dimensional reduction) to deal with the UV divergences.\footnote{
It is again easy to get the results for the usual dimensional regularization: then $\eta_{\r\s}\eta^{\r\s}=d$ and $\G^{(2)}_{\l\n}(q)\Big\vert^{\p'}$ gets an extra factor $2^{-\e/2}$.
} 
Indeed, before 
introducing the IR cutoff $\m$, the integrands must be brought into the ``standard form" according to the rules discussed in sect. 2.3: first introduce a Feynman parameter $x$ to rewrite the denominators ${1\over [p^2+m^2][(p+q)^2+m^2]} =\int_0^1 {\rm d}x {1\over [(p+xq)^2+R(x)]^2}$ where we define
\be\label{Rexp}
R(x)=m^2 + x(1-x) q^2 \quad , \quad R_0(x)=x(1-x) q^2 \ .
\ee
Then we shift the loop momentum from $p$ to $p'=p+xq$ which, of course,  is a well-defined operation only for convergent integrals, and this is why it is much more convenient to deal with dimensionally regularized integrals from the start. It is only then that we impose the IR cutoff $\m$. We get
\ba\label{diffa2}
{i\over 2}\,\G^{(2)}_{\l\n}(q)\Big\vert^{A'}
&=&-{1\over 2}\int_0^1 {\rm d}x\,
\Big\{\eta_{\r\s}\eta^{\r\s} 
\left(\eta_{\l\n} I_1(0) -2 I_{2,\l\n}(R_0) 
-{(1-2x)^2\over 2} q_\l q_\n I_2(R_0)\right)
\nonumber\\
&&
\hskip2.2cm +4\, (q_\l q_\n-q^2\eta_{\l\n}) I_2(R_0) \Big\} \ ,
\\
\label{diffgh2}
{i\over 2}\,\G^{(2)}_{\l\n}(q)\Big\vert^{\o'}
&=& \int_0^1 {\rm d}x\,
\left( \eta_{\l\n} I_1(0) -2 I_{2,\l\n}(R_0) 
-{(1-2x)^2\over 2} q_\l q_\n I_2(R_0)\right) \ ,
\\
\label{diffpsi2}
{i\over 2}\,\G^{(2)}_{\l\n}(q)\Big\vert^{\p'}
&=&
2\, \int_0^1 {\rm d}x\,
\left( \eta_{\l\n} I_1(R) -2I_{2,\l\n}(R) 
+2 (q_\l q_\n-q^2\eta_{\l\n}) x(1-x) I_2(R) \right) \ ,
\\
\label{diffphi2}
{i\over 2}\,\G^{(2)}_{\l\n}(q)\Big\vert^{\phi'}
&=&- \int_0^1 {\rm d}x\,
\left( \eta_{\l\n} I_1(m^2) -2 I_{2,\l\n}(R) -{(1-2x)^2\over 2} q_\l q_\n I_2(R)\right)\ ,
\ea
with the integrals $I_1$ and $I_2$ given in
(\ref{mLintegraldim}), (\ref{mLint2dim}) and (\ref{tensorintdim}).
The remaining integrals over the Feynman parameter $x$ are elementary. We found it convenient to introduce
\be\label{xirho} 
\xi={q^2\over \m^2} 
\quad , \quad
\rho={q^2\over \m^2+m^2} 
\quad , \quad
g(\xi)={1\over \sqrt{1+4/\xi}} 
\log{\sqrt{1+4/\xi}+1\over\sqrt{1+4/\xi}-1} \ ,
\ee
as well as the following functions
\ba\label{ffcts}
f_2(\xi)&=& {(2-\xi)(\xi+4)\over 12\, \xi^2}g(\xi)
-{1\over 3\xi} +{5\over 36}
= -{\xi\over 60}+{\xi^2\over 560} +{\cal O}(\xi^3) \ ,
\nonumber\\
f_3(\xi)&=& {g(\xi)\over \xi^2} 
-{1\over 2\xi} +{1\over 12}
\hskip2.5cm=\hskip3.mm{\xi\over 60}-{\xi^2\over 280} +{\cal O}(\xi^3)
\ ,\quad
\ea
which are such that $f_i(\xi)$ and $f_i(\rho)$ all vanish at $q^2=0$.
The various contributions to the  part of the Wilsonian one-loop effective action that is quadratic in $A$ then are
\ba\label{GoneloopqcontA}
\G_{\m,\rm 1-loop}\Big\vert_{A^2}^{A'}
&=&
c_d\, \int {{\rm d}^4 q\over (2\pi)^4} \Bigg\{ \left[ {5\over 6}\left({2\over\e}-\log\m^2\right) 
+10 f_2(\xi)-8f_3(\xi) \right] 
\tra \tilde F^{\rm lin}_{\l\s}(q)\tilde F_{\rm lin}^{\l\s}(-q)
\nonumber\\
&&\hskip2.4cm + \m^2\left[ 1+ 2\xi f_3(\xi)-{\xi\over 6}\right]
\tra \tilde A_\l(q) \tilde A^\l(-q) \Bigg\} \ ,
\\
\G_{\m,\rm 1-loop}\Big\vert_{A^2}^{\o'}
&=&
c_d\,\int {{\rm d}^4 q\over (2\pi)^4} \Bigg\{ \left[ {1\over 12}\left({2\over\e}-\log\m^2\right) 
+f_2(\xi) \right] 
\tra \tilde F^{\rm lin}_{\l\s}(q)\tilde F_{\rm lin}^{\l\s}(-q)
\nonumber\\
&&\hskip2.4cm -{\m^2\over 2} \left[ 1+ 2\xi f_3(\xi)-{\xi\over 6}\right]
\tra \tilde A_\l(q) \tilde A^\l(-q) \Bigg\} \ ,
\ea
and
\ba\label{Goneloopqcontp}
\G_{\m,\rm 1-loop}\Big\vert_{A^2}^{\p'}
&=&
2c_d\, \int {{\rm d}^4 q\over (2\pi)^4} \Bigg\{ \Bigg[ -{1\over 6}\left({2\over\e}-\log(\m^2+m^2)-{m^2\over \m^2+m^2}\right) 
-2 f_2(\r) +2 f_3(\r)
\nonumber\\
&&\hskip4.cm 
-{2m^2\over \m^2+m^2}f_3(\rho) \Bigg] \trR \tilde F^{\rm lin}_{\l\s}(q)\tilde F_{\rm lin}^{\l\s}(-q)
\nonumber\\
&&\hskip2.8cm -{1\over 2}{\m^4\over \m^2+m^2}\left[
1+ 2\r f_3(\r) -{\r\over 6}\right]
\trR \tilde A_\l(q) \tilde A^\l(-q) \Bigg\} \ ,
\\
\G_{\m,\rm 1-loop}\Big\vert_{A^2}^{\phi'}
&=&
c_d \int {{\rm d}^4 q\over (2\pi)^4} \Bigg\{ \Bigg[ -{1\over 12}\left({2\over\e}-\log(\m^2+m^2)-{m^2\over \m^2+m^2}\right)  -f_2(\rho)
\nonumber\\
&&\hskip4.cm
+{m^2\over \m^2+m^2}\left( {4+\r\over 2} f_3(\r) -{\r\over 24}\right)
\Bigg] 
\trR \tilde F^{\rm lin}_{\l\s}(q)\tilde F_{\rm lin}^{\l\s}(-q)
\nonumber\\
&&\hskip2.8cm +{1\over 2}{\m^4\over \m^2+m^2}\left[
1+ 2\r f_3(\r) -{\r\over 6}\right]
\trR \tilde A_\l(q) \tilde A^\l(-q) \Bigg\} \ ,
\ea
where 
\be\label{cd}
c_d= {1\over (4\pi)^{d/2}\G({d\over 2})} 
\ ,
\ee 
and $\tilde F^{\rm lin}_{\l\n}(q)=-i(q_\l \tilde A_\n(q)-q_\n \tilde A_\l(q))$ is the linearized part of the field strength. Again, for Majorana fermions the factor of $2$ is absent in $\G_{\m,\rm 1-loop}\Big\vert_{A^2}^{\p'}$. 

There are a few simple checks we can make: First we note that for $q^2=0$, and hence $\rho=\xi=0$, we have $f_i=0$ and $\tilde F_{\l\n}^{\rm lin}(q)=0$ and we recover the results obtained above for constant $A$, see eq. (\ref{quadraticA4}).
We can also compare the $q$-independent coefficients of 
$\int {{\rm d}^4 q\over (2\pi)^4} \tr \tilde F^{\rm lin}_{\l\s}(q)\tilde F_{\rm lin}^{\l\s}(-q)=\int {\rm d}^4 x\, 
\tr F^{\rm lin}_{\l\s}(x) F_{\rm lin}^{\l\s}(x)$ with the corresponding coefficients of $\tr F_{\l\s}F^{\l\s}$ in $\gamma_{\m,\rm 1-loop}\vert_{A^4}$ obtained above for constant $A$, see (\ref{quarticA'})-(\ref{quarticp'}). Modulo the replacement $\log\L \to {2\over\e}$ we get again perfect agreement.\footnote{Of course, in eq, (\ref{quarticp'}) there is an ambiguity concerning the $m^4I_4(m^2)$ coefficient of $\tr FF$ because we can change this coefficient by writing the non-gauge invariant term as $\tr A_\l A^\l A_\r A^\r$ instead of $\tr A_\l A_\r A^\l A^\r$. However since the non invariant terms cancel when adding $\g_{\m,\rm 1-loop}\vert^{\p'}_{A^4}$ and $\g_{\m,\rm 1-loop}\vert^{\phi'}_{A^4}$ one can compare the corresponding sums and one gets again perfect agreement. Due to this same ambiguity it is not useful to compare directly with (\ref{quarticA'dim})-(\ref{quarticphi'dim}).
}

\subsubsection{Supersymmetric gauge theories}

Next, we see that the different $\G_{\m,\rm 1-loop}\Big\vert_{A^2}$ contribute different non-gauge invariant terms\break\hfill
$\sim \tr \tilde A_\l(q) \tilde A^\l(-q)$, with $q$-dependent coefficients. For a general theory these non invariant terms do not cancel. However, there is again a perfect cancellation for the ${\cal N}=1$ vector and the ${\cal N}=1$ chiral multiplets:
\ba\label{N=1vectorq}
\G_{\m,\rm 1-loop}\Big\vert_{A^2}^{{\cal N}=1\ \rm vector}
&=&\hskip1.mm
{3\over 4}\ c_d\, \int {{\rm d}^4 q\over (2\pi)^4} 
\ \hat\g_\m(0,\xi)\
\tra \tilde F^{\rm lin}_{\l\s}(q)\tilde F_{\rm lin}^{\l\s}(-q) \ ,
\\
\nonumber\\
\label{N=1chiralq}
\G_{\m,\rm 1-loop}\Big\vert_{A^2}^{{\cal N}=1\ \rm chiral}
&=&\hskip-2.mm
-{1\over 4}\ c_d\, \int {{\rm d}^4 q\over (2\pi)^4} 
\ \hat\g_\m(m,\r)\ 
\trR \tilde F^{\rm lin}_{\l\s}(q)\tilde F_{\rm lin}^{\l\s}(-q) 
\ ,\ \quad
\ea
where we defined
\ba\label{N=1fct}
\hskip-5.mm
\hat\g_\m(m,\r)
&=&
\left({2\over\e}-\log(\m^2+m^2)\right) +12 f_2(\rho) -8 f_3(\rho) 
-{m^2\over\m^2+m^2}
\left(1 +2\r f_3(\rho)-{\r\over 6}\right)
\nonumber\\
&=&
\left({2\over\e}-\log(\m^2+m^2)\right) -{\r\over 3} 
+{\r^2\over 20}
-{m^2\over \m^2+m^2} \left(1 -{\r\over 6}+{\r^2\over 30}\right)
+{\cal O}(\r^3) \ .\ \quad
\ea
Thus, we see again that, in supersymmetric gauge theories, the terms of the Wilsonian effective action we computed indeed are gauge invariant.
Similarly for the ${\cal N}=2$ vector and hyper multiplets:
\ba\label{N=2vectorq}
\G_{\m,\rm 1-loop}\Big\vert_{A^2}^{{\cal N}=2\ \rm vector}
&=&\hskip1.mm
{1\over 2} \ c_d\, \int {{\rm d}^4 q\over (2\pi)^4} 
\ \hat\g_\m(0,\xi)\ 
\tra \tilde F^{\rm lin}_{\l\s}(q)\tilde F_{\rm lin}^{\l\s}(-q) 
\ ,
\\
\nonumber\\
\label{N=2hyperq}
\G_{\m,\rm 1-loop}\Big\vert_{A^2}^{{\cal N}=2\ \rm hyper}
&=&\hskip-2.mm
-{1\over 2} \ c_d\, \int {{\rm d}^4 q\over (2\pi)^4} 
\ \hat\g_\m(m,\r)\ 
\trR \tilde F^{\rm lin}_{\l\s}(q)\tilde F_{\rm lin}^{\l\s}(-q) 
\ .
\ea
Finally, for the ${\cal N}=4$ multiplet, one gets the sum of (\ref{N=2vectorq}) and (\ref{N=2hyperq}) with $m=0$ and $R$ taken to be the adjoint :
\be\label{N=4q}
\G_{\m,\rm 1-loop}\Big\vert_{A^2}^{{\cal N}=4}
= 0 \ ,
\ee
and we explicitly see (at one loop) that for ${\cal N}=4$ not only the Wilsonian gauge coupling is not  renormalized, but there also are no higher derivative terms in the quadratic part of the Wilsonian effective action. One can consider other one-loop finite theories like ${\cal N}=4$ broken to ${\cal N}=2$ by giving a mass $m$ to the hyper multiplet. It is actually known \cite{HSW} that ${\cal N}=2$ theories that are finite at one loop are finite to all orders in perturbation theory. Although being finite,  the one-loop contributions to the effective action for this theory are non-vanishing and e.g. the terms quadratic in the gauge field are 
\be\label{N=4toN=2}
\G_{\m,\rm 1-loop}\Big\vert_{A^2}^{{\cal N}=4\ \to \ {\cal N}=2}
={1\over 2} \ c_d\, \int {{\rm d}^4 q\over (2\pi)^4} \left[\hat\g_\m(0,\xi)-\hat\g_\m(m,\r)\right]\tra \tilde F^{\rm lin}_{\l\s}(q) \tilde F_{\rm lin}^{\l\s}(-q) \ .
\ee

\subsection{Higher-derivative Wilsonian couplings in susy gauge theories}

Let us write the first few terms in a derivative expansion of the Wilsonian effective action in a general ${\cal N}=1$ susy gauge theory with one vector multiplet and $n_i$ chiral multiplets with masses $m_i$ in representations $R_i$. We let $\tra t^a t^b=C_{\rm adj} \delta^{ab}$ as well as $\tr_{R_i} t^a t^b=C_{R_i}\delta^{ab}$. Then 
\be\label{derivexp}
\int {{\rm d}^4 q\over (2\pi)^4} \r_i^n \tr_{{\rm adj}/R_i} 
\tilde F^{\rm lin}_{\l\s}(q)\tilde F_{\rm lin}^{\l\s}(-q)
={C_{{\rm adj}/R_i}\over \left(\m^2+m_i^2\right)^n}  
\int {\rm d}^4 x\, 
(\del_{\n_1}\ldots \del_{\n_n} F^{\rm lin}_{a\l\s}(x)) 
\del^{\n_1}\ldots \del^{\n_n}F_{\rm lin}^{a\l\s}(x) \ ,
\ee
and similarly for $\xi^n$ with $m^2=0$.
If we assume that {\it all} terms in $\G_{\rm 1-loop}$ are gauge invariant, $F^{\rm lin}_{\l\s}$ must actually be the full $F_{\l\s}$ and the derivatives must be covariant derivatives. Thus we deduce from (\ref{N=1vectorq}), (\ref{N=1chiralq}) and (\ref{N=1fct}) that the terms involving two $F$'s and an arbitrary number of their derivatives\footnote{
As is well known, terms with different orderings of the covariant derivatives can be rewritten in the same form plus terms involving less derivatives and more than two $F$'s, see e.g. \cite{NBI}.
} 
in the Wilsonian effective action in an ${\cal N}=1$ susy gauge theory are
\be\label{YMWilsonian}
\G_\m= \int {\rm d}^4 x \left[ -{1\over 4 g_{(0)}^2(\m)} F^a_{\l\s} F_a^{\l\s}
+ \sum_{n=1}^\infty g_{(2n)}(\m) (D_{\n_1}\ldots D_{\n_n}F_{\l\s})^a (D^{\n_1}\ldots D^{\n_n}F^{\l\s})_a 
+ \ldots\right] \ ,
\ee
with the first few Wilsonian couplings given up to one-loop order (i.e. now including also the tree-level and counterterm contributions) by
\ba\label{wilsonianYMcoupl}
{1\over  g_{(0)}^2(\m)} 
&=& {1\over  g^2} +  {1\over (4\pi)^2} \left[3 C_{\rm adj}\, \log\m^2  
- \sum_i n_i C_{R_i} \, 
\left( \log(\m^2+m_i^2) +{m_i^2\over \m^2+m_i^2}\right)  + \hat c\right] \ ,
\nonumber\\
g_{(2)}(\m) &=&  - {1\over 12 (4\pi)^2} \left[
3 C_{\rm adj}\ {1\over \m^2}
- \sum_i n_i C_{R_i}\ {2\m^2+m_i^2\over 2(\m^2+ m_i^2)^2} \right] \ ,
\nonumber\\
g_{(4)}(\m) &=& {1\over 80 (4\pi)^2}\left[ 3 C_{\rm adj}\ {1\over\m^4}
- \sum_i n_i C_{R_i}\ {3\m^2+m_i^2\over 3(\m^2+ m_i^2)^3} \right] \ .
\ea
The precise value of the finite constant $\hat c$ depends on the choice of renormalization condition. Of course, the expansions of $f_2$ and $f_3$ immediately allow us to extract similarly {\it all} the one-loop Wilsonian couplings $g_{(2n)}$. Obviously also, there are many more terms in the Wilsonian action that we did not compute.

\vskip2.mm
\centerline{***}
\vskip2.mm

In this section, we have seen that the introduction of the IR cutoff $\m$ spoils the gauge invariance of the one-loop effective action, but when summing the contributions over full ${\cal N}=1$ susy multiplets the non-invariant terms cancel. This is reminiscent of anomalies where the contributions of individual chiral fermions to the effective action are not invariant but when summing over appropriate sets of fields the anomaly cancels. In this case there is a powerful theorem that no anomaly can occur at more than one loop. Of course, our non-gauge invariant terms have a structure that is very different from the topological character of ``ordinary anomalies". Nevertheless, it is tempting to speculate that one can prove a similar theorem also in the present case. 


\section{Conclusion}
\setcounter{equation}{0}

In this note we have given a detailed critical discussion of the properties of Wilsonian effective actions $\G_\m$. In particular, we have given a precise prescription how to implement the infrared cutoff $\m$ in any loop integral. At least at one loop, it is completely obvious that the full momentum integrals are reproduced when using $\G_\m$ as the action now with a UV cutoff $\m$. This prescription is manifestly Lorentz invariant and also preserves  global linear symmetries such as e.g. supersymmetry.  We have given a long discussion of the issue of gauge invariance of effective actions in general and in particular when using background field gauge. We have also discussed the approaches in the literature based on the exact renormalization group which are somewhat different in spirit. Our prescription of IR cutoff (as any similar prescription) breaks the gauge symmetry. Using our prescription, we have explicitly computed, at one loop, many terms of the Wilsonian effective action for general gauge theories involving bosonic and fermionic matter fields of arbitrary masses and in arbitrary representations, exhibiting  the non-gauge invariant (as well as the gauge invariant) terms. We have seen that for supersymmetric gauge theories all non-gauge invariant terms cancel within each supermultiplet. This is similar to the cancellation of anomalies for certain ``sets" of chiral fermions, and we have speculated that cancellation at one-loop is maybe enough to prove cancellation at any order. In any case, the cancellation provides strong evidence that in supersymmetric gauge theories one can indeed define a Lorentz, susy and gauge invariant Wilsonian action, which is the basic ingredient for the elegant proof of the non-renormalization theorems in \cite{SeibergNR}. We have given explicit formula in a general supersymmetric gauge theory for the one-loop Wilsonian couplings of various higher-derivative terms in the Wilsonian effective action.

\newpage

\appendixA{Appendix : Explicit IR cutoffs for $L$-loop integrals and a complete 2-loop computation}

{\bf \large A.1 The IR cutoff one-loop integrals}
\vskip2.mm

\noindent
First, let us indicate how one obtains the IR cutoff one-loop integrals in dimensional regularization, as given in eq. (\ref{mLint2dim}). After the obvious integration over the $S^{d-1}$ and setting $k_E^2=y$ they read
\be\label{INy}
I_N(R)={i\over (4\pi)^{d\over 2} \G({d\over 2})} \int_{\m^2}^\infty {\rm d}y\, {y^{{d\over 2}-1}\over (y+R)^N} \ .
\ee
For general $d=4-\e$ they are given in terms of incomplete Beta-functions. However, it is easy to extract the expansion for small $\e$. For $N=2$ one rewrites
\be\label{N=2int}
{1\over (y+R)^2}=-{R\over y(y+R)^2} - {R\over y^2 (y+R)} 
+ {1\over y^2} \ .
\ee
The first two terms lead to integrals that converge for $\e=0$ and hence can be evaluated directly at $\e=0$ where they are elementary, while the last term yields ${2\over \e}\m^{-\e}={2\over \e} - \log \m^2 +{\cal O}(\e)$. For $N=1$ one rewrites
\be\label{N=1int}
{1\over y+R}={R^2\over y^2 (y+R)} - {R\over y^2} + {1\over y} \ ,
\ee
with the first term giving a convergent integral when evaluated at $\e=0$, the second leading to $-{2\over \e}R+R\log\m^2 +{\cal O}(\e)$, and the last one, obtained by continuation from the region $\e>2$, yields $-\m^2$. Finally, for $N\ge 3$ the integrals are convergent and can be evaluated at $\e=0$. This leads to the integrals (\ref{mLint2dim}).

\vskip3.mm
\noindent
{\bf \large A.2 The prescription for $L$-loop integrals}
\vskip2.mm

\noindent
Now we will show how to extend our prescription for the IR cutoff to arbitrary $L$-loop diagrams. After introducing Feynman parameters $x_a$, an arbitrary $L$-loop diagram ${\cal G}$ involving $r$ propagators leads to a dimensionally regularized integral of the form 
\be\label{Imultiloop}
I_{\cal G}(p_s) 
= (r-1)! \left( \prod_{a=1}^r \int_0^1 {\rm d} x_a \right)
\delta\Big(\sum_{a=1}^r x_a -1\Big) 
\left( \prod_{i=1}^L {{\rm d}^d k_i\over (2\pi)^d} \right)
{\cal I}_{\cal G} \ ,
\ee
where
\be\label{Imultiint}
{\cal I}_{\cal G}=Q(k_i,p_s) \left[ \sum_{i,j=1}^L a_{ij}(x_a)
k_i\cdot k_j + 2 \sum_{i=1}^L k_i\cdot {\cal P}_i(x_a,p_s) + C(x_a,p_s,m_a)\right]^{-r} .
\ee
Here $k_i$ is the loop-momentum in the $i^{\rm th}$ loop, the $p_s$ are external momenta and the $m_a$ are the masses of the internal propagators. For generic values of the Feynman parameters, $[\ldots]$ is a non-degenerate, positive quadratic form in the $k_i$ which one can {\it diagonalize} by an orthogonal transformation $k_i=J_{ij}k'_j$. After appropriately shifting the $k'_i$ this assumes what we call the standard form
\be\label{Imultiloopstandard}
I_{\cal G}(p_s) 
= (r-1)! \left( \prod_{a=1}^r \int_0^1 {\rm d} x_a \right)
\delta\Big(\sum_{a=1}^r x_a -1\Big) 
\left( \prod_{i=1}^L {{\rm d}^d k'_i\over (2\pi)^d} \right)
{\cal I}'_{\cal G} \ ,
\ee
where now
\be\label{standardmulti}
{\cal I}'_{\cal G}=\tilde Q(k'_i,p_s) \left[ \sum_{i=1}^L A_{i}(x_a)
{k'_i}^2 + R(x_a,p_s,m_a)\right]^{-r}\ , 
\ee
the $A_i$ being the eigenvalues of $(a_{ij})$.
It is again on this standard form, after Wick rotating each $k'_i$, that one imposes the IR cutoff on each $(k'_i)_E^2\,$:
\be\label{kmutilde}
(k'_i)_E^2\ge \tilde\m^2_{\cal G} \ .
\ee
It should be clear that this $\tilde\m^2_{\cal G}$ does not necessarily need to be identical with $\m^2$ (which is the same for all diagrams), but may differ from it by a numerical factor depending on the topology of the diagram ${\cal G}$. Indeed, one may convince oneself that $\tilde\m^2_{\cal G}$ and $\m^2$ should be related by
\be\label{mumutilde}
\tilde\m^2_{\cal G}
= \m^2 \left[ (r-1)! \left( \prod_{a=1}^r \int_0^1 {\rm d} x_a \right)
\delta\Big(\sum_{a=1}^r x_a -1\Big) 
\sum_{i=1}^L A_i(x_a) \right]^{-1}  \ .
\ee
Clearly, for a one-loop diagram (where $A=1$) this gives back $\tilde\m^2_{\cal G}=\m^2$. For multi-loop diagrams that do {\it not} have several loops sharing a common propagator one has $\sum_i A_i(x_a)=1$ and this also gives $\tilde\m^2_{\cal G}=\m^2$. However, if several loops share a common propagator there is no reason why $\tilde\m^2_{\cal G}$ should equal $\m^2$ and (\ref{mumutilde}) provides the required correction factor. We will see an explicit example for both situations below and check that (\ref{mumutilde}) is indeed necessary for the consistency.

It is clear that this prescription for implementing the IR cutoff $\m$ is universal and unambiguous and that it preserves the various linear global symmetries. As repeatedly emphasized, an important consistency requirement is that, when computing correlation functions starting from the Wilsonian effective action and imposing a UV cutoff $\m$ using exactly the same prescription, the $\m$-dependence should cancel. This is obviously the case at one-loop. However, we were not able to provide a general proof beyond one loop. It might also be the case that a complete cancellation requires some further refinement of our prescription for higher loop diagrams. Of course, the difficulty is due to multi-loop diagrams containing different loops sharing a common propagator. It is similar to the complications encountered in the BPHZ renormalization program  when dealing with overlapping UV divergences. In the latter case the parametric representation can be useful to simplify this problem \cite{Zuber}, and maybe these techniques could be implemented here as well.

\vskip3.mm
\noindent
{\bf \large A.3 A complete two-loop computation}
\vskip2.mm

\begin{figure}[h]
\centering
\includegraphics[width=0.9\textwidth]{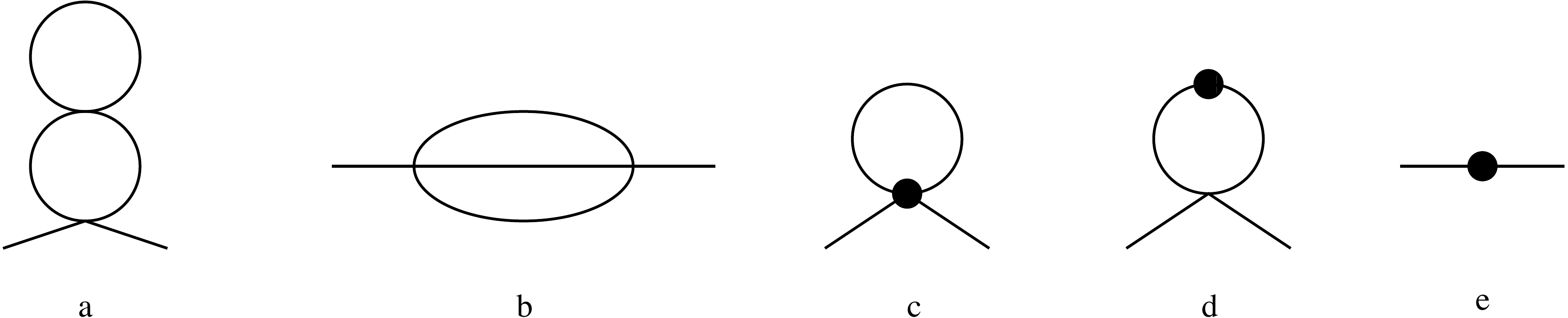}\\
\caption[]{Shown are the ${\cal O}(g^2)$ contributions to $\G^{(2)}_\m$: there are two two-loop diagrams (a) and (b) as well as two one-loop diagrams (c) and (d) involving counterterms, and the ${\cal O}(g^2)$ wave function renormalization counterterm (e).} \label{twoloop}
\end{figure}

\noindent
To see how our prescription for multi-loop integrals works in practice, we will compute $\G_\m^{(2)}$ up to order $g^2$ in scalar $\vf^4$-theory. In particular, this involves computing the two two-loop diagrams shown in Fig. \ref{twoloop}, using our IR-cutoff $\m$. 
\vskip2.mm

We will first compute the simpler diagram (a) shown on the left of Fig. \ref{twoloop}. The relevant integral for this diagram is $p$-independent and is simply
\be\label{I2a}
I_{(a)}^{(2)} = 
\int {{\rm d}^d k\over (2\pi)^d}{{\rm d}^d l\over (2\pi)^d}
{1\over [k^2+m^2]^2\ [l^2+m^2]} \ .
\ee
This factorizes as a product of two one-loop integrals, and one clearly should impose the IR cutoff for each loop separately as $k_E^2\ge \m^2$ and $l_E^2\ge \m^2$ which simply gives
\be\label{I2amu}
I_{(a)\,\m}^{(2)} = I_2(m^2)\, I_1(m^2) \ ,
\ee
with $I_1$ and $I_2$ given in (\ref{mLint2dim}). Let us alternatively apply our general prescription explained above and check that it leads to the same result:
\be\label{I2amup}
I_{(a)}^{(2)}= 2 \int_0^1 {\rm d}x\, {\rm d}y\, {\rm d}z\, 
\delta(x+y+z-1) 
\int {{\rm d}^d k\over (2\pi)^d}{{\rm d}^d l\over (2\pi)^d}\ {\cal I}_{(a)} \ ,
\ee
with
\be\label{IIa}
{\cal I}_{(a)}=\left[ x\, l^2 +(y+z)\, k^2 +m^2\right]^{-3}
\equiv \left[ A\, l^2 + B\, k^2 +m^2\right]^{-3} \ .
\ee
This already is in the standard form and, according to (\ref{kmutilde}), the IR cutoff is $l_E^2\ge \mt^2$ and $k_E^2\ge\mt^2$. We have $A+B=1$ and thus (\ref{mumutilde}) gives $\mt^2=\m^2$. Thus, in the present case, our general prescription reduces to the obvious IR cutoff $k_E^2\ge \m^2$ and $l_E^2\ge \m^2$ for each loop, and we get just (\ref{I2amu}).

For $m\ne 0$, (\ref{I2amu}) contains ${1\over \e^2}$ poles. The result is somewhat simpler for $m=0$, where only $I_2$ has a ${1\over \e}$ pole while  $I_1(0)=-{i\over (4\pi)^2}\m^2\left[ 1+\left( 2-\g+\log4\pi-\log\m^2\right){\e\over 2}+{\cal O}(\e^2)\right]$ is finite. We then get
\be\label{Iamumzero}
I^{(2)}_{(a)\, \m} = {1\over (4\pi)^4} \left(
{1\over \e} +{3\over 2} -\g +\log 4\pi -\log\m^2 \right) 2\m^2 
\quad , \qquad {\rm for}\ m=0 \ .
\ee
\vskip2.mm

Now we turn to the other two-loop diagram (b) shown in Fig. \ref{twoloop}, which is somewhat more complicated.
Labelling the loop-momenta by $k$, $l$ and letting $p$ be the external momentum, the relevant integral is
\ba\label{twoloopint}
I^{(2)}_{(b)}(p)&=&
\int {{\rm d}^d k\over (2\pi)^d}{{\rm d}^d l\over (2\pi)^d}
{1\over [(p+k+l)^2+m^2][l^2+m^2][k^2+m^2]} 
\nonumber\\
&=& 2 \int_0^1 {\rm d}x\, {\rm d}y\, {\rm d}z\, 
\delta(x+y+z-1) \tilde I(x,y,z)\ ,
\ea
with
\be\label{Ixyz}
\tilde I(x,y,z)=\int {{\rm d}^d k\over (2\pi)^d}{{\rm d}^d l\over (2\pi)^d}
[x(p+k+l)^2+y l^2 +z k^2 +m^2]^{-3} \ .
\ee
The quadratic form in $k$ and $l$ appearing in the denominator is diagonalized by the orthogonal transformation
\be\label{orthogonal}
\begin{pmatrix}k\\l\\\end{pmatrix}
=\begin{pmatrix}\cos\theta&\sin\theta\\
-\sin\theta&\cos\theta\\\end{pmatrix}
\begin{pmatrix}q\\r\\\end{pmatrix}
\quad , \quad 
\cos 2\theta = \sqrt{(y-z)^2\over (y-z)^2+4x^2} \ .
\ee
One then shifts $q$ and $r$ to bring the integrand in the standard form. After some algebra one gets
\be\label{Ixyzstandard}
\tilde I(x,y,z)=\int {{\rm d}^d q\over (2\pi)^d}{{\rm d}^d r\over (2\pi)^d}
[A\, q^2 + B\, r^2 + R]^{-3} \ ,
\ee
where $A$, $B$ and $R$ depend on $x$, $y$ and $z$ and are given by
\ba\label{ABR}
A&=& x+{y+z\over 2}-{1\over 2}\, {\rm sgn}(y-z)\sqrt{(y-z)^2+4x^2} \ ,
\nonumber\\
B&=& x+{y+z\over 2}+{1\over 2}\, {\rm sgn}(y-z)\sqrt{(y-z)^2+4x^2}\ ,
\nonumber\\
\nonumber\\
R&=& m^2 + {x y z\over xy+xz+yz}\, p^2 \ .
\ea
It is on this standard form (\ref{Ixyzstandard}), after Wick rotating each of the two integrations, that we impose the IR cutoff: $s\equiv q_E^2\ge\mt^2$ and $t\equiv r_E^2\ge\mt^2$. Thus
\ba\label{Ixyzmu}
\tilde I_\m(x,y,z)&=&-{1\over (4\pi)^d \G({d\over 2})^2} 
\int_{\mt^2}^\infty {\rm d}s\,s^{{d\over 2}-1}
\int_{\mt^2}^\infty {\rm d}t\,t^{{d\over 2}-1}
[A s + B t + R]^{-3} 
\nonumber\\
&=&-{(AB)^{-{d\over 2}}\over (4\pi)^d \G({d\over 2})^2}
\int_{\mt^2 A}^\infty {\rm d}s\,
\int_{\mt^2 B}^\infty {\rm d}t\ {(st)^{{d\over 2}-1}\over (s+t+R)^3}
\ .
\ea

Even if this integral were UV convergent in $d=4$ one should not set $d=4$ at this stage since the integrals over the Feynman parameters $x$, $y$ and $z$ typically are divergent for $d=4$, and $d\ne 4$ also regularizes these integrals. To perform the latter one lets $x=1-u$, $y=uv$ and $z=u(1-v)$ so that
\be\label{Iuv}
I^{(2)}_{(b)\,\m} (p)= 2 \int_0^1 {\rm d}v\, \int_0^1 {\rm d}u\, u\
\tilde I_\m(1-u,uv,u(1-v)) \ .
\ee
With this parametrization we have 
\ba\label{ABRuv}
AB=u(1-\s(v)u) \quad &,& \quad A+B=2-u \ ,
\nonumber\\
\nonumber\\
R=m^2+{u(1-u)(1-\s(v))\over (1-\s(v)u)}\, p^2 
\quad &,& \quad \s(v)=1-v(1-v) \ .
\ea
Then eq. (\ref{mumutilde}) gives
\be\label{mumutildeb}
\mt^2\ =\m^2\ \left[2\int_0^1 {\rm d}v\, \int_0^1 {\rm d}u\, u\ (A+B)\right]^{-1} \ =\ {3\over 4}\, \m^2 \ .
\ee
Divergences arise in (\ref{Iuv}) either as $u\to 0$, or as $u\to 1$ and $\s(v)\to 1$.
To see this, consider first the case $\m=0$ where the integral (\ref{Ixyzmu}) is easily evaluated:
\be\label{Ixyzmu=0}
\tilde I_{\m=0}
=-{\G(3-d)\over 2(4\pi)^d} (A B)^{-{d\over 2}} R^{d-3}
\ .
\ee
For $m^2\ne 0$ the corresponding integrand in (\ref{Iuv}) is singular as $u^{-1+{\e\over 2}}$ for $u\to 0$ and as\break\hfill $(1-\s(v)u)^{-2+{\e\over 2}}$ for $u\to 1,\ \s(v)\to 1$. Both lead to a ${1\over \e}$ pole. Together with $\G(3-d)$ this produces a double pole. However, we will be mostly interested in the case $m=0$. Then the integrand is less singular since it behaves as $u^{\e\over 2}$ for $u\to 0$ and as $(1-\s(v))\,(1-\s(v)u)^{-3+{\e\over 2}}$ for $u\to 1, \s(v)\to 1$, giving a finite result even at $\e=0$ (apart from the ${1\over \e}$ pole from $\G(3-d)$). Thus, expanding the integrand to first order in $\e$, one obtains quite easily 
\ba\label{Imu=0}
I^{(2)}_{(b)\,\m=0}(p)\Big\vert_{m=0}&=&-{\G(3-d)\over (4\pi)^d}\, (p^2)^{d-3}\, \int_0^1 u{\rm d}u \int_0^1 {\rm d}v 
{\Big[ u^2(1-u) v(1-v)\Big]^{d-3}
\over \Big[u [1-u(1-v(1-v))]\Big]^{{3d\over 2}-3}}
\nonumber\\
&=&{1\over 2(4\pi)^4} \left\{ \left( {1\over \e}+{13\over 4}-\g+\log4\pi\right)p^2
- p^2 \log p^2  +{\cal O}(\e) \right\}\ ,
\ea
exhibiting a divergent $p^2$ piece and a finite {\it non-local} term $p^2 \log p^2$, as expected in a massless theory in the absence of an IR-cutoff.

Let us now return to $\tilde I_\m$ in the presence of the IR cutoff $\m$. We will now {\it restrict ourself to the massless case} which is really the interesting one. To compute this double integral (\ref{Ixyzmu}) one has to resort to the same type of rewriting of the integrand as we already used to evaluate the one-loop diagrams, see (\ref{N=2int}) and (\ref{N=1int}), namely $(s+t+R)^{-3}= \left[(s+t+R)^{-3}-g(s,t,R)\right] + g(s,t,R)$ with $g(s,t,R)$ chosen such that the first term $[\ldots]$ yields a UV convergent integral for $\e=0$ and such that the second term can be explicitly evaluated even for $\e\ne 0$ . We will take
\be\label{gstR}
g(s,t,R)= {1\over (s+t)^3} - {3 R\over (s+t)^4} \ .
\ee
This corresponds to separating off the first two terms in a Taylor expansion in $R$.
Accordingly we have 
$\tilde I_\m =\tilde I_{\m,1} + \tilde I_{\m,2}$ with
\ba\label{Im12}
\tilde I_{\m,1}&=& -{(AB)^{-{d\over 2}}\over (4\pi)^d \G({d\over 2})^2}
\int_{\mt^2 A}^\infty {\rm d}s\,
\int_{\mt^2 B}^\infty {\rm d}t\ (st)^{{d\over 2}-1}
\left[ {1\over (s+t+R)^3}-g(s,t,R)\right] \ ,
\nonumber\\
\tilde I_{\m,2}&=& -{(AB)^{-{d\over 2}}\over (4\pi)^d \G({d\over 2})^2}
\int_{\mt^2 A}^\infty {\rm d}s\,
\int_{\mt^2 B}^\infty {\rm d}t\ (st)^{{d\over 2}-1}\,
g(s,t,R) \ .
\ea
Let us first compute $\tilde I_{\m,1}$. The integrals over $s$ and $t$ are convergent for $d<5$ and $\tilde I_{\m,1}$ can be expanded in powers of $\e=4-d$: $\tilde I_{\m,1}= \tilde I_{\m,1}^{(0)} + \e\, \tilde I_{\m,1}^{(1)} + \ldots$. As explained above, the integrations over the ``Feynman parameters" $u$ and $v$ potentially lead to singularities, so that one might have to keep the pre-factor $(AB)^{-{d\over 2}}$ as such. However, one can convince oneself\footnote{
This will be obviously the case for $\tilde I_{\m,1}^{(0)}\equiv \tilde I_{\m,1}\vert_{d=4}$ given below. We have also computed $\tilde I_{\m,1}^{(1)}$  and checked explicitly that its integrals over $u$ and $v$ are non-singular, too.  
}
that no such singularities are generated when integrating 
$\tilde I_{\m,1}^{(0)}$ or $\tilde I_{\m,1}^{(1)}$, and hence we can set $\e=0$ and obtain
\be\label{Im1d4}
\tilde I_{\m,1}\Big\vert_{d=4}
=- {(AB)^{-2} R\over 2 (4\pi)^4}
\left\{ \log\left( 1+{R\over \mt^2(A+B)}\right) 
+{AB\over (A+B)^2}\, {R\over \mt^2(A+B)+R} \right\} \ .
\ee
Next, the integral $\tilde I_{\m,2}$ decomposes in an obvious way as
\ba\label{Jnint}
\tilde I_{\m,2} &=& 
-{(AB)^{-{d\over 2}} \over (4\pi)^d \left(\G({d\over 2})\right)^2}
\left( J^{(3)} - 3 R J^{(4)} \right) \ ,
\nonumber\\
J^{(n)}&=& \int_{\mt^2 A}^\infty {\rm d}s\,
\int_{\mt^2 B}^\infty {\rm d}t\ {(st)^{{d\over 2}-1}\over (s+t)^n} \ .
\ea
The $J^{(n)}$ can be expressed in terms of hypergeometric functions which one can then expand in $\e$. Alternatively, with some care, one can obtain the leading and subleading terms in an expansion in $\e$ directly as elementary integrals. The result is\footnote{
As usual in dimensional regularization, quadratically divergent integrals like $J^{(3)}$ or the one-loop integral $I_1$ are defined by continuation from their region of convergence which is $d<2$. This is why $J^{(3)}$, just as $I_1(0)$, has no ${1\over \e}$ pole. We have scaled out a factor $\mt^{2-2\e}$, resp. $\mt^{-2\e}$, from $J^{(3)}$, resp. $J^{(4)}$, before expanding in $\e$.
}
\ba\label{J34}
J^{(3)} &=& 
{(\mt^2)^{1-\e}\over 2} \left\{ 
{AB\over A+B} - (A+B)+ {\e\over 2} \left[ (A+B)\Big(\log(A+B)-1\Big) +{A^2\log A+B^2\log B\over A+B}\right]
\right\}  \ ,
\nonumber\\
J^{(4)} &=&  {(\mt^2)^{-\e}\over 6}
\left\{ {1\over \e} +{A B\over (A+B)^2} - \log(A+B)\right\}  \ .
\ea

It remains to perform the integrations over the ``Feynman parameters" $u$ and $v$. For $\tilde I_{\m,1}$ we get
\be\label{pF}
2 \int_0^1 {\rm d}v\,\int_0^1 {\rm d}u\,u\ 
\tilde I_{\m,1}\Big\vert_{d=4}
=-{1\over (4\pi)^4}\ p^2\, F({p^2\over \mt^2}) \ ,
\ee
with
\ba\label{Fxi}
F(\xi)&=&\int_0^1 {\rm d}v\,\int_0^1 {\rm d}u\,
{(1-u)(1-\s)\over (1-\s u)^3} \Bigg\{
\log\left( 1+{u(1-u)(1-\s)\over (2-u)(1-\s u)} \xi\right)
\nonumber\\
&& \hskip5.5cm
+ {u^2(1-u)(1-\s)\over (2-u)^3} \xi\left(1+ {u(1-u)(1-\s)\over (2-u)(1-\s u)}\xi\right)^{-1} \Bigg\}
\nonumber\\
&\simeq& 0.02687\ \xi - 0.00091\ \xi^2 + 0.00005\ \xi^3 + {\cal O}(\xi^4) \ ,
\ea
with  $\s\equiv\s(v)=1-v(1-v)$. When performing the integral of $\tilde I_{\m,2}$, the part coming from $J^{(4)}$ gives no further singularity and one can expand $(AB)^{-{d\over 2}}=(AB)^{-2} \big(1+{\e\over 2}\log(AB)\, \big)$. The integrals then are elementary.\footnote{
We have $\int_0^1 {\rm d}v\,\int_0^1 {\rm d}u\,u\ (AB)^{-2} R={p^2\over 2}$ and $\int_0^1 {\rm d}v\,\int_0^1 {\rm d}u\,u\ (AB)^{-2} R \left[{1\over 2} \log(AB) -\log(A+B) +{AB\over (A+B)^2}\right]= -(\log 2+{1\over 4} ) {p^2\over 2}$.

}
On the other hand, the part coming from $J^{(3)}$ generates a ${1\over \e}$ pole plus a finite piece. To get the finite part correctly requires some care.\footnote{
First, the part in $J^{(3)}$ that multiplies $\e$, only contributes at the singularities of $u\,(AB)^{-2+\e/2}=u^{-1+\e/2} (1-\s u)^{-2+\e/2}$ which are at $\{u=0\}$ and $\{u=1,\ \s(v)=1\}$. This leads to 
$\int_0^1 {\rm d}v\,\int_0^1 {\rm d}u\,u\ (AB)^{-2+\e/2}\ {\e\over 2}
\left[ (A+B)\Big(\log(A+B)-1\Big) +{A^2\log A+B^2\log B\over A+B}\right]=4\log2-4$. Then, in the integral $\int_0^1 {\rm d}v\,\int_0^1 {\rm d}u\,u\ (AB)^{-2+\e/2}\left( {AB\over (A+B)} -(A+B)\right)$ one extracts the singular parts which are $\int_0^1 {\rm d}v\,\int_0^1 {\rm d}u\,u^{-1+\e/2} (-2) =-{4\over \e}$ and $\int_0^1 {\rm d}v\,\int_0^1 {\rm d}u\, (1-\s(v)u)^{-2+\e/2} (-1)=-{4\over \e}-2$, the remainder then is non-singular and can be evaluated at $\e=0$ yielding $2{\cal C}-2$, ${\cal C}$ being Catalan's costant. Thus $\int_0^1 {\rm d}v\,\int_0^1 {\rm d}u\,u\ (AB)^{-2+\e/2} J^{(3)}={1\over 2}(\mt^2)^{1-\e} \left(-{8\over \e}+2{\cal C}-8+4\log2\right)$.
} 
We get
\be\label{I2mp}
2\int_0^1 {\rm d}v\,\int_0^1 {\rm d}u\,u\  \tilde I_{\m,2}
={1\over (4\pi)^4}\left[ 
\left({1\over\e}+a+\log{3\over 4}-\log\mt^2\right)8\mt^2 +\left({1\over\e}+b+\log{3\over 4}-\log\mt^2\right){p^2\over 2}
\right] \ ,
\ee
with
\ba\label{bvalue}
a&=&2-\g-{{\cal C}\over 4}+\log2\pi+{1\over 2}\log2-\log{3\over 4}\ ,
\nonumber\\
b&=&{3\over 4}-\g+\log 2\pi-\log{3\over 4} \ ,
\ea
where ${\cal C}\simeq 0.915966$ is Catalan's constant. Thus, collecting all the pieces, and substituting $\mt^2={3\over 4} \m^2$ (cf. (\ref{mumutildeb})), we finally get
\be\label{I2final}
I_{(b)\,\m}^{(2)}(p)={1\over (4\pi)^4}\left[ 
\left({1\over\e}+a-\log\m^2\right)6\m^2 +\left({1\over\e}+b-\log\m^2\right){p^2\over 2}
-p^2 F\left({4 p^2\over 3 \m^2}\right) \right] \ .
\ee
Note that this corresponds to a {\it local} contribution to the Wilsonian action $\G_\m$, since $F\left({4 p^2\over 3 \m^2}\right)$ and hence $I_{(b)\, \m}^{(2)}(p)$ can be expanded in a series in $p^2$, as long as $\vert{p^2\over \m^2}\vert$ is not too large. In fact, it is easy to see numerically that $F(\xi)$ has a singularity at $\xi \simeq -11.5$, so the expansion is possible as long as $\vert{p^2\over \m^2}\vert \lesssim 8.6$. If $p^2$ is too large, or $\m^2$ too small one sees again the cross-over to the non-local behavior, just as in the one-loop examples studied in section 2.2.

It is interesting to take the limit $\m\to 0$ and see how (\ref{I2final}) matches (\ref{Imu=0}). In this limit one needs the asymptotics of $F(\xi)$ for $\xi\to\infty$. It is then easy to perform the integrals in $F(\xi)$ and one gets
\be\label{Flimit}
F(\xi)
\ \begin{matrix} {}\\ \sim \\ {}^{\xi\to \infty} \end{matrix}\ 
{1\over 2}\left( \log\xi -{5\over 2}-\log 2\right) \ .
\ee
We see that the $p^2\log\m^2$ term then cancels and we get
\be\label{I2mnlimit}
I_{(b)\, \m}^{(2)}(p)
\ \begin{matrix} {}\\ \sim \\ {}^{\m\to 0} \end{matrix}\ 
{1\over 2(4\pi)^4}\left[ 
\left({1\over\e}+b+{5\over 2}+\log 2 +\log{3\over 4}\right) p^2 -p^2 \log p^2
\right] \ .
\ee
Given the value (\ref{bvalue}) of $b$, this matches (\ref{Imu=0}) exactly.

Let us now give the contributions of these two two-loop diagrams to the Wilsonian $\G_\m^{(2)}$.  Each integral gets multiplied by a $(-g)^2$ for the two vertices and a $(-i)^2$ for the two loops. More importantly, diagram $(a)$ comes with a symmetry factor ${1\over 4}$ while diagram $(b)$ has a symmetry factor ${1\over 6}$. Thus from (\ref{Iamumzero}) and (\ref{I2final})
\ba\label{Gtwoloop}
\hskip-9.mm \G_\m^{(2)}(p)\Big\vert_{\rm 2-loop}\hskip-2.mm  
&=&\hskip-2.mm 
- g^2 \left( {1\over 4}\ I_{(a)\, \m}^{(2)}(p)
+ {1\over 6}\ I_{(b)\, \m}^{(2)}(p) \right)
\nonumber\\
&=&\hskip-2.mm 
- {g^2\over (4\pi)^4} \left[
\left( {3\over 2\e} -{3\over 2} \log\m^2 +\tilde a \right) \m^2
+  \left({1\over\e}+b-\log\m^2\right) {p^2\over 12}
-{p^2\over 6} F\left({4 p^2\over 3 \m^2}\right) \right] ,\ \ 
\ea
where $\tilde a=a+{3\over 4}-{\g\over 2}+{1\over 2}\log4\pi$, and $a$ and $b$ are given in (\ref{bvalue}). Note that the coefficient ${3\over 2}$ of ${\m^2\over \e}$ originates as ${1\over 4}\times 2+ {1\over 6} \times 6$ where the last $6$ arose from converting the $8\mt^2$ to $6\m^2$.

There are two more ${\cal O}(g^2)$ contributions to $\G_\m^{(2)}$ which are one-loop and involve counterterms. The first corresponds to the one-loop diagram (c) of Fig. \ref{twoloop} involving the $\vf^4$ counterterm needed to make the $\G_\m^{(4)}$ finite at one loop. As discussed in section 2.4.1, this counterterm is $\int {\rm d}^4 x\ \left(-{1\over 4!}\right) {3 g^2\over (4\pi)^2 }\left({1\over \e} + {c_0\over 2}\right) \vf^4$, where the value of the finite constant $c_0$ depends on the renormalization condition. 
Thus the contribution of diagram (c) to $\G_\m^{(2)}(p)$ is (still for $m=0$)
\ba\label{Gloopcounter}
\G_\m^{(2)}(p)\Big\vert_{\rm 1-loop/counterterm}
&=&{1\over 2}\left(-{3 g^2\over (4\pi)^2 }\right)
\left({1\over \e} +{c_0\over 2}\right)
(-i) I_1(0) 
\nonumber\\
&=&-{g^2\over (4\pi)^4 } \left( -{3\over 2\e} +{3\over 4}\log\m^2
+\hat c_0 \right) \m^2
\ \ ,
\ea
where $\hat c_0=-{3\over 4}(c_0+2-\g +\log 4\pi)$. Diagram (d) involves the ${\cal O}(g)$ (mass renormalization) counterterm. The latter is $\sim m^2$ and, since we are restricting ourselves here to $m=0$, it is absent and diagram (d) gives no contribution. Finally, there is the wave-function renormalization counterterm of diagram (e) which just gives a 
\be\label{wfct}
\G_\m^{(2)}(p)\Big\vert_{\rm counterterm}={g^2\over (4\pi)^4 }\,  
\left( {1\over \e} +b-\hat b\right) {p^2\over 12}\ , 
\ee
designed to cancel the ${p^2\over \e}$ term in (\ref{Gtwoloop}). Again, the value of $\hat b$ depends on the renormalization conditions.

Adding (\ref{Gtwoloop}), (\ref{Gloopcounter}) and (\ref{wfct}), we finally obtain the full ${\cal O}(g^2)$ contribution to $\G_\m^{(2)}$:
\be\label{G2fulltl}
\G_\m^{(2)}(p)\Big\vert_{g^2} 
= - {g^2\over (4\pi)^4} \left[
\left( \hat a  - {3\over 4} \log\m^2 \right) \m^2
+  \left(\hat b-\log\m^2\right)\, {p^2\over 12}
-{p^2\over 6}\, F\left({4 p^2\over 3 \m^2}\right) \right] \ ,\ \ 
\ee
where 
$\hat a=\tilde a+\hat c_0$ is given by
\be\label{ahat}
\hat a = -{3\over 4}(c_0+\g-\log4\pi) +{5\over 4}
-{{\cal C}\over 4}-{1\over 2}\log{9\over 8} \ ,
\ee
and depends, via $c_0$, on the renormalization conditions.
We observe that the terms $\sim {\m^2\over \e}$ have cancelled, as they indeed should. Clearly, the presence of such a $\m$-dependent diverging term would have been a disaster since we are not allowed to add a $\m$-dependent counterterm to the action. The cancellation of these terms constitutes a non-trivial consistency check of our two-loop computation and in particular of the relation (\ref{mumutilde}) giving $\mt$ in terms of $\m$. Furthermore, loosely speaking, the $\m^2\log\m^2$ and $p^2\log\m^2$ terms are the remnants in the renormalized $\G^{(2)}_\m$ of the UV divergences of the loop diagrams. It is then not too difficult to see that, when computing the 1PI $\G^{(2)}\vert_{g^2}\,$, the various $\m^2\log\m^2$ and $p^2\log\m^2$ terms cancel.



\begin{thebibliography}{99}

\bibitem{Pol}
J.~Polchinski,
{\it Renormalization and effective Lagrangians},
Nucl.\ Phys.\ B {\bf 231} (1984) 269.

\bibitem{supergraphnonren}
M.~T.~Grisaru, W.~Siegel and M.~Rocek,
{\it Improved methods for supergraphs},
Nucl.\ Phys.\  B {\bf 159} (1979) 429.

\bibitem{SeibergNR}
N.~Seiberg, {\it Naturalness versus supersymmetric nonrenormalization theorems},
Phys.\ Lett.\  B {\bf 318} (1993) 469,
[arXiv:hep-ph/9309335].


\bibitem{ZJ}
J. Zinn-Justin, {\it Renormalization of gauge theories}, in {\it Trends in elementary particle theory, International summer institute on theoretical physics in Bonn 1974}, Springer-Verlag Berlin 1975.

\bibitem{BV}
I.~A.~Batalin and G.~A.~Vilkovisky,
{\it Gauge algebra and quantization},
Phys.\ Lett.\  B {\bf 102} (1981) 27;
{\it Closure of the gauge algebra, generalized Lie equations and Feynman rules},
Nucl.\ Phys.\  B {\bf 234} (1984) 106.

\bibitem{West} 
P. West, {\it Introduction to supersymmetry and supergravity}, 
chap. 17, World Scientific 1990.

\bibitem{BB}
I.~L.~Buchbinder, E.~I.~Buchbinder, S.~M.~Kuzenko and B.~A.~Ovrut,
{\it The background field method for N = 2 super Yang-Mills theories in harmonic superspace},
Phys.\ Lett.\  B {\bf 417} (1998) 61,
[arXiv:hep-th/9704214];
I.~L.~Buchbinder and B.~A.~Ovrut,
{\it Background field method and structure of effective action in N=2  super Yang-Mills theories},
[arXiv:hep-th/9802156].

\bibitem{Burgess}
C.~P.~Burgess,
{\it Introduction to effective field theory},
[arXiv:hep-th/0701053].

\bibitem{SV} 
M.A. Shifman and A.I. Vainshtein, {\it Solution of the anomaly puzzle
in susy gauge theories and the Wilson operator expansion},
Nucl. Phys. B {\bf 277} (1986) 456, {\it On holomorphic dependence and 
infrared effects in supersymmetric gauge theories}, 
Nucl. Phys. B {\bf 359} (1991) 571.

\bibitem{Warr}
B. Warr,
{\it Renormalization of gauge theories using effective Lagrangians. 1}, Annals Phys.\  {\bf 183} (1988) 1;
{\it Renormalization of gauge theories using effective Lagrangians. 2}, Annals Phys.\  {\bf 183} (1988) 59.
  
\bibitem{Wetterich}
J.~Berges, N.~Tetradis and C.~Wetterich,
{\it Non-perturbative renormalization flow in quantum field theory and statistical physics},
Phys.\ Rept.\  {\bf 363} (2002) 223,
[arXiv:hep-ph/0005122].

\bibitem{Peskin}
M.~E.~Peskin and D.~V.~Schroeder,
{\it An introduction to quantum field theory},
Addison-Wesley 1995.

\bibitem{Siegel}
W. Siegel, 
{\it Inconsistency of supersymmetric dimensional regularization},
Phys. Lett. {\bf 94B} (1980) 37.

\bibitem{AGW}  
L.~Alvarez-Gaum\'e and E.~Witten,
{\it Gravitational anomalies},
Nucl.\ Phys.\  B {\bf 234} (1984) 269.

\bibitem{Bechi}
C. Bechi, 
{\it On the construction of renormalized gauge theories using renormalization group techniques}, 
Elementary Particle, Field Theory and Statistical Mechanics, Eds. M. Bonini, G. Marchesini and E. Onofri, Parma University 1993,
[arXiv:hep-th/9607188].

\bibitem{Sonoda}
H. Sonoda,
{\it On the construction of QED using ERG},
[arXiv:hep-th/0703167];
Y. Igarashi, K. Itoh, H. Sonoda,
{\it Quantum master equation for QED in exact renormalization group},
[arXiv:0704.2349].

\bibitem{Kopper}
G. Keller and C. Kopper,
{\it Perturbative renormalzation of QED via flow equations},
Phys. Lett. B {\bf 273} (1991) 323;
{\it Renormalizability proof for QED based on flow equations},
Commun.\ Math.\ Phys.\  {\bf 176} (1996) 193.

\bibitem{Bonini}  
M. Bonini, M. D'Attanasio, G. Marchesini,
{\it Ward identities and Wilson renormalisation group for QED},
Nucl. Phys. B {\bf 418} (1994) 81.  
  
\bibitem{Ellwanger}
U.~Ellwanger,
{\it Flow equations and BRS invariance for Yang-Mills theories},
Phys.\ Lett.\  B {\bf 335} (1994) 364,
[arXiv:hep-th/9402077].
    
\bibitem{Rosten}
O. Rosten, T. Morris, S. Arnone,
{\it Manifestly gauge invariant QED},
JHEP {\bf 0510} (2005) 115,
[arXiv:hep-th/0505169];
{\it A generalised manifestly gauge invariant exact renormalisation group for SU(N) Yang-Mills},
Eur. Phys. J. {\bf C 50} 467-504 (2007),
[arXiv:hep-th/0507154];
T. Morris and O. Rosten,
{\it Manifestly gauge invariant QCD}, J.Phys. A39 (2006) 11657-11681,
[arXiv:hep-th/0606189].

\bibitem{Luscher}
M.~L\"uscher,
{\it Abelian chiral gauge theories on the lattice with exact gauge invariance},
Nucl.\ Phys.\  B {\bf 549} (1999) 295,
[arXiv:hep-lat/9811032].
  
\bibitem{KL}
V.~Kaplunovsky and J.~Louis,
{\it Field dependent gauge couplings in locally supersymmetric effective quantum field theories},
Nucl.\ Phys.\  B {\bf 422} (1994) 57,
[arXiv:hep-th/9402005].

\bibitem{Weinbook}
S. Weinberg, {\it The quantum theory of fields}, vol.II, chap. 17.5, Cambridge University Press 1996.

\bibitem{HSW}
P.~S.~Howe, K.~S.~Stelle and P.~C.~West,
{\it A class of finite four-dimensional supersymmetric field theories},
Phys. Lett. B {\bf 124} (1983) 55.

\bibitem{NBI}
E.~A.~Bergshoeff, A.~Bilal, M.~de Roo and A.~Sevrin,
{\it Supersymmetric non-abelian Born-Infeld revisited},
JHEP {\bf 0107} (2001) 029, [arXiv:hep-th/0105274];
A.~Bilal,
{\it Higher-derivative corrections to the non-abelian Born-Infeld action},
Nucl.\ Phys.\  B {\bf 618} (2001) 21, [arXiv:hep-th/0106062].
  
\bibitem{Zuber}
M.-C. Berg\`ere and J.-B. Zuber, 
{\it Renormalization of Feynman amplitudes and parametric integral
representation},
Commun.\ Math.\ Phys.\  {\bf 35} (1974) 113.


\end{thebibliography}
\end{document}